\documentclass[journal]{IEEEtran}
\usepackage{amsmath,amsfonts}
\usepackage{algorithmic}
\usepackage{algorithm}
\usepackage{array}
\usepackage[caption=false,font=scriptsize,labelfont=sf,textfont=sf]{subfig}
\usepackage{textcomp}
\usepackage{stfloats}
\usepackage{url}
\usepackage{verbatim}
\usepackage{graphicx}
\usepackage{cite}
\usepackage{booktabs}
\usepackage{bm}
\usepackage{makecell}
\usepackage[table]{xcolor}
\definecolor{lightblue}{RGB}{0,176,240}
\usepackage{pifont}
\usepackage{arydshln}
\usepackage{multirow}
\begin{document}
\makeatletter
\renewcommand\@IEEEBIOskipN{-0.5    \baselineskip} 
\makeatother

\title{Surveillance Facial Image Quality Assessment: A Multi-dimensional Dataset and Lightweight Model}

\author{Yanwei Jiang, Wei Sun, Yingjie Zhou, Xiangyang Zhu, Yuqin Cao, Jun Jia, Yunhao Li, \\ Sijing Wu, Dandan Zhu, Xingkuo Min, Guangtao Zhai,~\IEEEmembership{Fellow,~IEEE}

\IEEEcompsocitemizethanks{\IEEEcompsocthanksitem 
\\ \indent Yanwei Jiang, Yingjie Zhou, Yuqin Cao, Jun Jia, Yunhao Li, Sijing Wu, Xiongkuo Min, and Guangtao Zhai are with the Institute of Image Communication and Network Engineering, Shanghai Jiao Tong University, 200240, Shanghai, China. (e-mail:\{jiang-yan-wei, zyj2000, caoyuqin, jiajun0302, lyhsjtu, wusijing, minxiongkuo, zhaiguangtao\}@sjtu.edu.cn).
\\ \indent Wei Sun is with School of Communication and Electronic Engineering, East China Normal University, 200241, Shanghai, China. (e-mail: wsun@cee.ecnu.edu.cn).
\\ \indent Dandan Zhu is with the School of Computer Science and Technology, East China Normal University, 200333, Shanghai, China (e-mail: ddzhu@mail.ecnu.edu.cn).
\\ \indent Xiangyang Zhu is with Shanghai AI Lab, 200232, Shanghai, China. (e-mail: zhuxiangyang@pjlab.org.cn).
\\ \indent Corresponding authors: Wei Sun (wsun@cee.ecnu.edu.cn).
\\ \indent This work was supported by the National Natural Science Foundation of China (62301316, 62225112).
\\ \indent Copyright © 2026 IEEE. Personal use of this material is permitted.
However, permission to use this material for any other purposes must be obtained
from the IEEE by sending an email to pubs-permissions@ieee.org.
 \protect} }

\markboth{IEEE Transactions on Circuits and Systems for Video Technology}{Jiang \MakeLowercase{\textit{et al.}}: Surveillance Facial Image Quality Assessment: A Multi-dimensional Dataset and Lightweight Model}

\maketitle

\begin{abstract}
Surveillance facial images are often captured under unconstrained conditions, resulting in severe quality degradation due to factors such as low resolution, motion blur, occlusion, and poor lighting. Although recent face restoration techniques applied to surveillance cameras can significantly enhance visual quality, they often compromise fidelity (\emph{i.e.}, identity-preserving features), which directly conflicts with the primary objective of surveillance images---reliable identity verification. Existing facial image quality assessment (FIQA) predominantly focus on either visual quality or recognition-oriented evaluation, thereby failing to jointly address visual quality and fidelity, which are critical for surveillance applications. To bridge this gap, we propose the first comprehensive study on surveillance facial image quality assessment (SFIQA), targeting the unique challenges inherent to surveillance scenarios. Specifically, we first construct \textit{SFIQA-Bench}, a multi-dimensional quality assessment benchmark for surveillance facial images, which consists of $\bm{5,004}$ surveillance facial images captured by three widely deployed surveillance cameras in real-world scenarios. A subjective experiment is conducted to collect six dimensional quality ratings, including noise, sharpness, colorfulness, contrast, fidelity and overall quality, covering the key aspects of SFIQA. Furthermore, we propose \textit{SFIQA-Assessor}, a lightweight multi-task FIQA model that jointly exploits complementary facial views through cross-view feature interaction, and employs learnable task tokens to guide the unified regression of multiple quality dimensions. The experiment results on the proposed dataset show that our method achieves the best performance compared with the state-of-the-art general image quality assessment (IQA) and FIQA methods, validating its effectiveness for real-world surveillance applications. 

\end{abstract}


\section{Introduction}

In the digital era, public surveillance systems have rapidly proliferated, becoming a cornerstone of modern intelligent infrastructures. Surveillance facial images (SFIs), as the core visual input for these systems, play an indispensable role in public security, identity verification, and access control scenarios~\cite{choung2024acceptance}. However, unlike portrait images captured in controlled environments, SFIs are often acquired under highly unconstrained conditions, which can lead to severe quality degradation caused by factors such as low resolution, motion blur, occlusion, non-ideal shooting angles, and poor or uneven lighting. These degradations pose significant challenges for both human visual inspection and automated face recognition systems when processing SFIs.

\begin{figure}[!t]
\setlength{\abovecaptionskip}{0pt}
  \centering
  \includegraphics[width=0.95\linewidth]{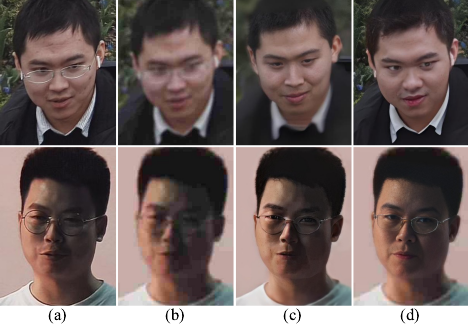}
  \caption{(a) High-quality reference facial images. (b) Corresponding degraded versions generated by applying JPEG compression and blur. (c) and (d) Restored images produced by two representative face restoration methods~\cite{yang2021gan,yue2024difface}. While the restorations enhance perceptual quality, they may introduce hallucinated textures or pseudo-structures, potentially compromising identity fidelity. }
  \label{fig:restoration}
\end{figure}

In light of these challenges, recent advances in face restoration~\cite{yang2021gan,yue2024difface,zhou2022towards,wang2021towards} have been applied to surveillance imagery with the aim of improving visual quality. While these techniques can improve the perceptual appearance of degraded images to a certain extent, these methods may also introduce pseudo-structures or hallucinated details, thus compromising face fidelity (see Fig.~\ref{fig:restoration}), which directly undermines the primary objective of surveillance imaging: reliable identity verification. This concern underscores the urgent need for a reliable and comprehensive quality assessment model that can effectively evaluate both perceptual quality and face fidelity in surveillance facial images, ensuring trustworthy face-related applications in real-world scenarios.

Existing facial image quality assessment (FIQA) studies primarily focus on two aspects. The first is perceptual FIQA, such as TransFQA~\cite{liu2024assessing}, and DSL-FIQA~\cite{chen2024dsl}, which aim to assess the perceived visual quality of facial images based on human subjective ratings. The facial images in these studies are mainly sourced from the Internet or social media platforms and are annotated with quality scores provided by human viewers. However, these datasets lack facial images captured in real-world surveillance scenarios and do not account for the impact of facial enhancement techniques on face fidelity, thereby limiting their applicability in surveillance contexts. 

\begin{figure}[!t]
\setlength{\abovecaptionskip}{-10pt}
  \centering
  \includegraphics[width=0.9\linewidth]{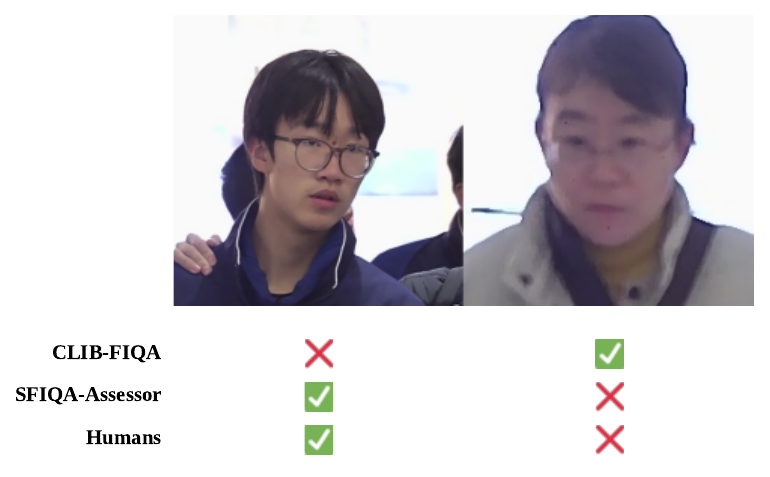}
  \caption{Illustration of the misalignment between recognition quality and perceptual quality. Checkmarks indicate much higher quality score than crossmarks. CLIB-FIQA scores reflect recognition quality, while SFIQA-Bench scores represent perceptual quality. Compared to CLIB-FIQA, the proposed SFIQA-Assessor produces scores more consistent with human judgments.}
  \label{fig:percept_rec}
\end{figure}

The second is recognition-oriented FIQA, represented by methods such as FaceQNet~\cite{hernandez2019faceqnet}, SER-FIQ~\cite{terhorst2020ser}, and CLIB-FIQA~\cite{ou2024clib}, which aim to assess the recognition quality of facial images based on their utility in face recognition systems. These methods typically use facial images from face recognition datasets as the evaluation targets. In this setting, quality labels are usually derived from face recognition models rather than human judgments. For example, the matching score between a test image and its corresponding enrolled image is often used as a pseudo ground-truth indicator of quality. However, due to the inherent robustness of modern face recognition systems to various distortions, there is often a misalignment between recognition quality and perceptual quality. That is, an image with high recognition quality does not necessarily exhibit high perceptual quality, as illustrated in Fig.~\ref{fig:percept_rec}. Furthermore, existing recognition-oriented FIQA studies also rarely consider the impact of facial restoration techniques on face fidelity. These limitations render recognition-based quality indicators less suitable for surveillance applications.

To bridge these gaps, we present the first comprehensive study on surveillance facial image quality assessment, which systematically addresses   the challenges of assessing both perceptual quality and face fidelity in real-world surveillance scenarios. To achieve this goal, we first construct \textbf{SFIQA-Bench}, a multi-dimensional quality assessment benchmark for SFIs. In contrast to previous FIQA studies that primarily collect facial images from the Internet, where imaging techniques may be outdated and the capture conditions differ significantly from those in surveillance contexts, our dataset comprises real-world surveillance images captured using three widely deployed surveillance cameras. These cameras are equipped with integrated face enhancement modules that are automatically triggered in poor capture environments. In total, we collect 5,004 surveillance facial images across three representative scenarios: indoor person surveillance, outdoor pedestrian surveillance, and intelligent transportation system (ITS) surveillance for vehicle occupants, covering diverse environmental conditions such as variations in ambient lighting, capture angles, etc. This ensures broad generalizability and strong practical utility of the benchmark.

Since SFIs are often captured under challenging conditions, such as low lighting, extreme pitch angles, motion, etc., they typically exhibit diverse and complex distortions. To systematically investigate their perceptual quality, we evaluate SFIs across four dimensions: \textbf{noise}, \textbf{sharpness}, \textbf{colorfulness}, and \textbf{contrast}, which allows us to identify the specific causes of visual degradation and supports explainable perceptual quality assessment. To address the unnatural artifacts introduced by face enhancement techniques, we introduce a \textbf{fidelity} dimension to evaluate whether an image contains unrealistic, unnatural, or hallucinated facial content. We also include an \textbf{overall quality} dimension to analyze the relative importance of each factor and its correlation with overall quality. To obtain reliable annotations, we organized a large-scale subjective experiment involving $100$ participants. Each image was rated by at least $25$ subjects across all six dimensions under controlled viewing conditions, ensuring high-quality and consistent annotations.

Furthermore, we propose \textbf{SFIQA-Assessor}, a lightweight deep neural network (DNN)-based model specifically designed for real-time, multi-dimensional quality assessment of SFIs. The model takes as input three complementary facial views: the original image containing both the face and background, a cropped face image focusing on the facial region, and an eyes-and-mouth region emphasizing key facial features. To integrate these views, we propose a \textbf{facial-aware quality feature encoder} that extracts multi-scale features from the three views and fuses them via a lightweight cross-view attention module to efficiently capture inter-view dependencies and form a unified quality representation. Subsequently, we design a \textbf{task-aware quality decoder} that incorporates both task self-attention and cross-attention modules. A set of learnable task tokens is first refined via task self-attention to model inter-task relationships, and then used as queries in cross-attention to extract task-specific features from the unified representation. Finally, separate \textbf{regression heads} are used to map each task-specific feature to its corresponding quality score. Experimental results demonstrate that SFIQA-Assessor consistently outperforms state-of-the-art methods across all quality dimensions while maintaining extremely low computational complexity, validating its practical value for real-world surveillance applications.

Our contributions are summarized as follows:
\begin{itemize}
\item We construct SFIQA-Bench, the first multi-dimensional quality assessment benchmark tailored for surveillance facial images. It contains 5,004 images captured under three representative real-world surveillance scenarios using three widely deployed camera models, ensuring broad generalizability and practical utility.

\item We conduct a comprehensive subjective experiment on SFIQA-Bench to collect human ratings across six quality dimensions: noise, sharpness, colorfulness, contrast, fidelity, and overall quality, providing interpretable and fine-grained labels for both perceptual quaity and face fidelity evaluation of surveillance facial images.

\item We propose SFIQA-Assessor, a lightweight FIQA model that extracts and fuses multi-scale features from multiple facial views to assess six quality dimensions, achieving state-of-the-art performance with real-time efficiency.
\end{itemize}

\section{Related Work}

\subsection{Image Quality Assessment}

Image quality assessment (IQA)\footnote{In this paper, we focus on no-reference IQA (also known as blind IQA) methods, as the task of FIQA typically falls under this category.} has been extensively studied over the years, with numerous datasets~\cite{sheikh2006statistical,larson2010most,ciancio2010no,virtanen2014cid2013,ponomarenko2015image,ghadiyaram2015massive,ying2020patches,lin2019kadid,hosu2020koniq,fang2020perceptual,zhang2023synergetic} and models~\cite{mittal2012making,mittal2012no,moorthy2011blind,kang2014convolutional,bosse2017deep,ke2021musiq,sun2023blind,chen2024topiq,wang2023exploring,yang2022maniqa,wu2024q,zhang2023blind,madhusudana2022image,zhang2023synergetic,gao2023blind,zhou2024multitask,zhou2024blind,sun2022graphiqa, zhou2025better,liu2022liqa,wang2024blind,zhou2021no,xu2020blind,zhou2020tensor,zhou2019dual,chen2017blind,liu2025multi,cao2025multi} proposed. Early IQA datasets primarily focus on synthetic distortions such as JPEG compression, Gaussian noise, blur, etc. Popular datasets like LIVE~\cite{sheikh2006statistical}, CSIQ~\cite{larson2010most}, and TID2013~\cite{ponomarenko2015image} typically consist of a limited set of high-quality reference images, each distorted using several common synthetic distortions, resulting in a total of a few hundred to several thousand distorted images. The KADID-10k~\cite{lin2019kadid} dataset significantly expanded the scale of synthetic IQA benchmarks by incorporating 81 pristine images and 25 types of distortions, resulting in over 10,000 distorted samples. In recent years, the focus of IQA research has shifted toward in-the-wild images, which are directly captured by cameras or collected from the Internet, and exhibit authentic distortions arising from real-world acquisition and transmission conditions. Representative in-the-wild IQA datasets include BID~\cite{ciancio2010no}, CLIVE~\cite{ghadiyaram2015massive}, KonIQ-10k~\cite{hosu2020koniq}, SPAQ~\cite{fang2020perceptual}, and FLIVE~\cite{ying2020patches}, with dataset sizes ranging from several hundred to over ten thousand images. Detailed information about these datasets is provided in Table~\ref{tab:Datasets}.

IQA models can be broadly categorized into two main groups: handcrafted feature-based methods and DNN-based approaches. Handcrafted feature-based methods usually rely on natural scene statistics (NSS)~\cite{mittal2012making,mittal2012no,moorthy2011blind,tang2014noise} or other manually designed features such as texture~\cite{liu2009no,narvekar2011no,sebastian2012gray}, edge~\cite{ferzli2009no,bahrami2014fast,vu2009s3}, or free energy cues~\cite{zhai2011psychovisual}. Classical handcrafted feature-based methods, such as DIIVINE~\cite{moorthy2011blind}, BRISQUE~\cite{mittal2012no}, NIQE~\cite{mittal2012making}, etc., are mainly designed for synthetically distorted images. However, they tend to generalize poorly to in-the-wild scenarios due to their limited modeling capacity. In contrast, DNN-based approaches leverage the powerful representation learning abilities of deep neural networks to extract quality-aware features, achieving better performance on both synthetic and authentic distortions. Early DNN-based models~\cite{kang2014convolutional,bosse2017deep,zhang2020blind} typically employ shallow convolutional neural networks (CNNs) that operate on image patches to learn local quality representations. More recent works~\cite{ke2021musiq,sun2023blind,chen2024topiq,gao2024no,wang2023exploring,yang2022maniqa,wu2024q,zhang2023blind,min2025exploring} incorporate advanced architectures such as multi-scale feature extractors~\cite{ke2021musiq,sun2023blind,gao2024no}, attention mechanisms~\cite{yang2022maniqa,chen2024topiq,min2025exploring}, semantic priors from CLIP variants~\cite{wang2023exploring,zhang2023blind}, and large multimodal models (LMMs)~\cite{wu2024q} to enhance robustness and accuracy. To further improve generalization, several studies explore multi-dataset training~\cite{zhang2023blind,gao2023blind} or adopt unsupervised learning strategies~\cite{madhusudana2022image}, enabling the use of large-scale unlabeled data and improving model adaptability across diverse conditions.

\subsection{Facial Image Quality Assessment}

Existing FIQA studies primarily fall into two categories: perceptual FIQA and recognition-oriented FIQA.

Perceptual FIQA focuses on the perceived visual quality of facial images, aligning more closely with conventional IQA research. Su \textit{et al.}~\cite{su2023going}  introduce GFIQA-20K, a FIQA dataset containing Internet-sourced 20,000 facial images annotated with subjective quality scores, and propose a generative prior-guided IQA model for facial quality prediction. Jo \textit{et al.}~\cite{jo2023ifqa} propose IFQA, a novel interpretable face quality assessment method that uses a generative-adversarial framework to simulate face restoration and provide pixel-level quality scores aligned with human perception. IFQA emphasizes key facial regions such as the eyes, nose, and mouth, outperforming traditional general-purpose and face-specific quality metrics across various architectures and challenging scenarios. Chahine \textit{et al.}~\cite{chahine2023image} present a portrait IQA dataset comprising 5,116 images captured in 50 different scenes, with expert pairwise annotations for detail, exposure, and overall quality. They further propose SEM-HyperIQA, a semantics-aware model adapted to scene-specific quality scales. Sun \textit{et al.}~\cite{sun2024dual} propose a dual-branch network for portrait IQA, leveraging separate Swin Transformer backbones for full-image and facial-region feature extraction, optimized via a learning-to-rank strategy. Liu \textit{et al.}~\cite{liu2024assessing} construct a FIQA dataset containing both in-the-wild and synthetically distorted facial images, and propose TransFQA, a two-branch Transformer model that captures global, facial, and component-level contexts as well as distortion-aware cues to improve prediction accuracy. Chen \textit{et al.}~\cite{chen2024dsl} introduce CGFIQA-40K, a balanced and diverse FIQA dataset comprising 40,000 images, designed to mitigate skin tone and gender biases. They also propose DSL-FIQA, which employs a dual-set degradation representation learning strategy that leverages both real and synthetic distortions to enhance generalization, and incorporates facial landmark-guided attention to focus on perceptually important regions.

Recognition-oriented FIQA aims to estimate the utility of a facial image for recognition tasks by predicting a quality score that correlates with recognition performance. FaceQnet~\cite{hernandez2019faceqnet,hernandez2020biometric} formulates this task as a supervised regression problem by using similarity scores from a pre-trained face recognition model as pseudo labels. To avoid reliance on ground-truth quality annotations, SER-FIQ~\cite{terhorst2020ser} introduces an unsupervised method that performs multiple stochastic forward passes with dropout in a recognition network and measures the embedding consistency as a proxy for quality. SDD-FIQA~\cite{ou2021sdd} adopts a self-supervised strategy by computing the Wasserstein distance between intra-class and inter-class similarity distributions, capturing how well an image distinguishes between identities. CLIB-FIQA~\cite{ou2024clib} further leverages a vision-language model (\textit{e.g.}, CLIP~\cite{radford2021learning}) to generate pseudo quality labels based on interpretable visual degradation factors such as blur and occlusion, and employs a confidence calibration mechanism to mitigate the influence of noisy supervision.

\begin{table*}[!t]
  \caption{Summary of Awesome IQA/FIQA Datasets}
  \label{tab:Datasets}
  \resizebox{\textwidth}{!}{
  \begin{tabular}{ccccccccc}
    \toprule
    Dataset & Task & Image Type& Source & Evaluation Dimensions & \# Images & Resolution & Label & Distortion\\
    \midrule
    LIVE~\cite{sheikh2006statistical}   & IQA & Generic & Professional Camera & Overall quality & 779 &  $\leq 768\times 512 $ & DMOS & Synthetic  \\
    CSIQ~\cite{larson2010most}          & IQA & Generic & Professional Camera & Overall quality & 866&  $512\times 512$ & MOS &  Synthetic \\
    TID2013~\cite{ponomarenko2015image} & IQA & Generic & Professional Camera & Overall quality & 3,000 &  $512\times 384 $& MOS & Synthetic  \\
    KADID-10k~\cite{lin2019kadid} & IQA  & Generic& Professional Camera & Overall quality & 10,125 &  $512\times 384 $ & DMOS & Synthetic \\
    \hdashline
    BID~\cite{ciancio2010no}            & IQA & Generic & Professional Camera & Overall quality & 6,585 &  $ 2272\times 1704 $ & MOS & Authentic  \\
    CLIVE~\cite{ghadiyaram2015massive}  & IQA & Generic & Smartphone & Overall quality & 1,162 &  $1024\times 768$& MOS & Authentic \\
    KonIQ-10k~\cite{hosu2020koniq} & IQA & Generic& Internet & Overall quality & 10,073 &  $1024\times 768 $ & MOS & Authentic \\
    SPAQ~\cite{fang2020perceptual} & IQA & Generic& Smartphone & \makecell{Brightness, Colorfulness, Contrast\\Noisiness, Sharpness, Overall Quality} & 11,125 &  $\leq 768\times 512 $& MOS & Authentic\\
    FLIVE~\cite{ying2020patches}        & IQA & Generic & Internet & Overall quality & 40,000 &  $640 \times 640$ & MOS& Authentic  \\
    \hdashline
    FERET~\cite{phillips2000feret} &  FIQA & Face& Professional Camera & Recognition quality & 14,126&  $256\times 384 $ & ID & Authentic \\
    LFW~\cite{huang2008labeled} & FIQA & Face& Internet & Recognition quality & 13,233 &  $250\times 250 $& ID & Authentic \\
    CASIA-WebFace~\cite{yi2014learning} & FIQA & Face& Internet & Recognition quality & 500,000 & - & ID &Authentic \\
    VGGFace2~\cite{cao2018vggface2} & FIQA & Face& Internet & Recognition quality & 3.31M & - & ID & Authentic\\
    \hdashline
    GFIQA-20K~\cite{su2023going} & FIQA & Face& Internet & Overall quality & 20,000 &  $512\times 512 $& MOS & Authentic\\
    PIQ23~\cite{chahine2023image} & FIQA & Portrait & Smartphone & Exposure, Detail, Overall quality & 5,116 &  $ 1900\times 2400 $& MOS & Authentic\\
    CGFIQA-40K~\cite{chen2024dsl} & FIQA & Face& Internet & Overall quality & 39,312 &  $512 \times 512 $& MOS & Authentic\\
    TransFQA-Dataset~\cite{liu2024assessing} & FIQA & Face & Internet & Overall quality & 42,125 & $\geq 512 \times 512$ & MOS & Hybrid \\
    \hdashline
    SFIQA-Bench & FIQA & Face & \makecell{Surveillance Camera} & \makecell{Noise, Sharpness, Colorfulness \\ Contrast, Fidelity, Overall quality} & 5,004&  $\leq 768\times 768 $ & MOS & Authentic\\
    \bottomrule
  \end{tabular}
  }

\end{table*}

\section{SFIQA-Bench}
In this section, we provide a detailed introduction to our surveillance facial image quality assessment benchmark, SFIQA-Bench, including the image collection process, the subjective quality evaluation, and benchmark analysis.

\subsection{Data Collection}
\begin{figure}[!t]
 \setlength{\abovecaptionskip}{0pt}
  \centering
  \includegraphics[width=\linewidth]{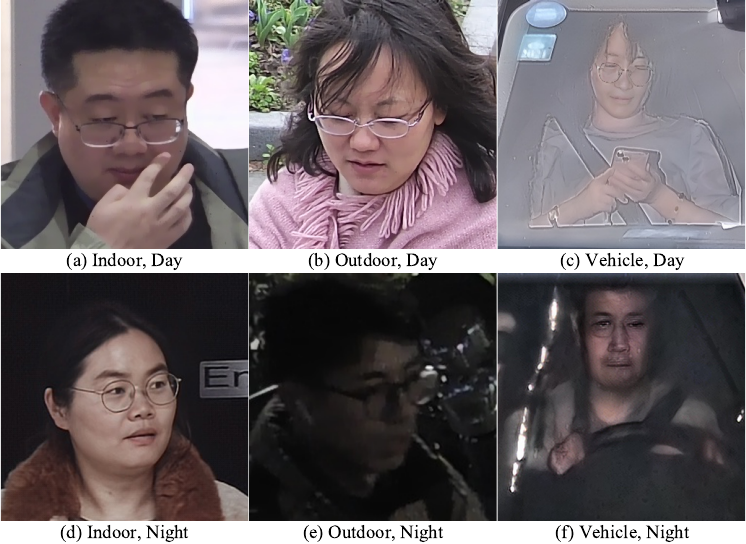}
  \caption{Representative examples of real-world surveillance facial images in SFIQA-Bench. The images cover six typical scenarios: indoor, outdoor, and in-vehicle surveillance captured during both daytime and nighttime.}
  \label{fig:scene_composite}
\end{figure}

We collect real-world surveillance images for SFIQA-Bench to capture authentic distortions arising from diverse environments and camera devices. Three representative scenarios are included: indoor person surveillance with simplified backgrounds under controlled lighting (on/off), outdoor pedestrian surveillance with complex backgrounds, and Intelligent Transportation Systems (ITS) surveillance for vehicle occupants---the latter two encompassing both daytime and nighttime conditions to reflect real-world lighting variations. Three distinct devices were utilized: the Huawei M2241-10-QLI-E2-B dual-lens camera, the HikVision DS-2CD7C888MWD dual-lens camera, and the HikVision DS-2CD7V887MWD four-lens surveillance camera. Each device operated in its built-in intelligent mode, automatically capturing facial images as individuals passed by, and then saving only the facial regions. Additionally, when detecting poor environmental conditions, the cameras activated an automatic facial enhancement mode to improve visual quality. Fig.~\ref{fig:scene_composite} illustrates example SFIs captured under various surveillance scenarios.
The cameras were mounted at heights between 2.5 and 3.5 meters, with pitch angles ranging from 15° to 30°. For indoor and outdoor scenes, cameras were installed in 6 office locations and 12 outdoor positions within a university campus. For the ITS scenario, surveillance images were collected in collaboration with the traffic police department. The image capturing process lasted two months, resulting in a total of 11,164 raw facial images. Subsequently, we conducted manual review and data cleaning. Images exhibiting excessively large yaw angles (greater than 60°) were removed, as were faces smaller than $96\times96$ pixels. Images with significant accidental occlusions, such as those obscured by hands, body parts of other individuals, or static objects like leaves, were also eliminated. Additionally, excessively similar images depicting the same individual were excluded. Regarding face masks, we chose not to exclude masked faces, as they have become commonplace in daily life.

\begin{figure}[!t]
 \setlength{\abovecaptionskip}{-5pt}
  \centering
  \includegraphics[width=\linewidth]{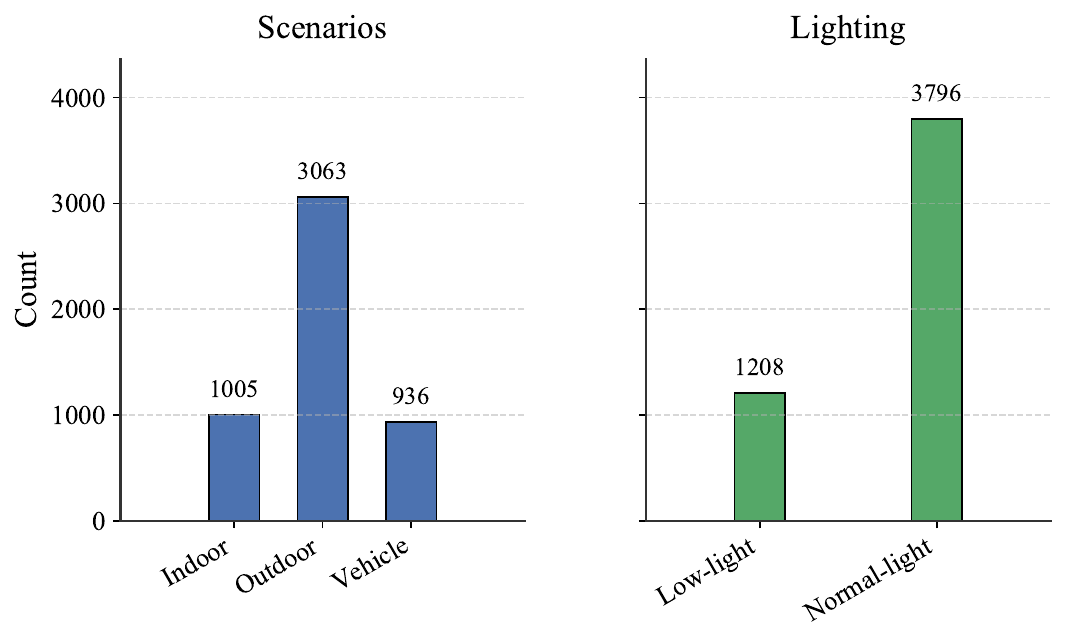}
  \caption{Distributions of some attributes in SFIQA-Bench.}
  \label{fig:attribute}
\end{figure}

\begin{figure}[!t]
\setlength{\abovecaptionskip}{-5pt}
  \centering
  \includegraphics[width=0.8\linewidth]{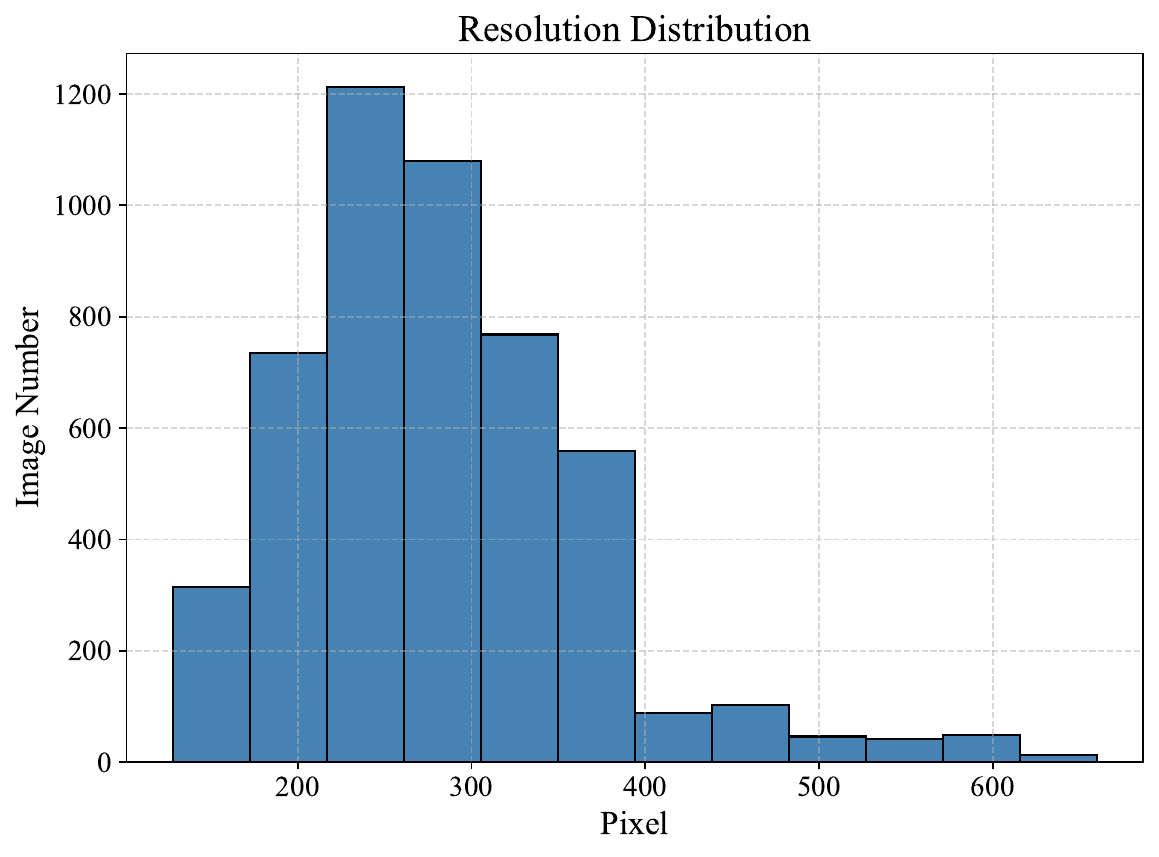}
  \caption{Histogram of image resolutions in SFIQA-Bench. All images are square, so only one side of the resolution is shown.}
  \label{fig:resolution}
\end{figure}

Following this cleaning process, the final dataset contains 5,004 surveillance facial images. Figs.~\ref{fig:attribute} and~\ref{fig:resolution} illustrate several key distributions, including scenario types (indoor, outdoor, and ITS vehicle scenes), lighting conditions (low-light and normal-light), image resolution histograms. These distributions show that our SFIQA-Bench encompasses diverse surveillance facial images consistent with real-world capturing conditions.


\subsection{Subjective Experiments}
Following the collection of SFIs, we carry out a rigorous human subjective evaluation to obtain quality scores for each image.

\subsubsection{Evaluation Dimensions}
We evaluate the quality of SFIs across multiple dimensions, as a single overall score is insufficient to capture the diverse and complex distortions present in such images. For instance, lighting conditions in surveillance settings vary widely, ranging from complete darkness (0 lux) to strong sunlight (over 10,000 lux). These variations result in large differences in image contrast. In low-light environments, infrared light is often used to reveal facial textures. However, combining infrared with RGB images can cause color shifts and produce misaligned or ghost-like textures. Motion blur is also common when the subject is moving. In addition, built-in facial enhancement modules may create unnatural or hallucinated facial features, reducing the fidelity of the facial image.

To comprehensively evaluate the common distortions, we define six quality dimensions: \textbf{noise}, \textbf{sharpness}, \textbf{colorfulness}, \textbf{contrast}, \textbf{fidelity}, and \textbf{overall quality}. The first four dimensions reflect perceptual quality attributes, while fidelity captures the authenticity and realism of the facial content. Overall quality represents a holistic assessment that considers all these quality aspects. The definitions of each dimension are detailed as follows:

\begin{itemize}
    \item \textbf{Noise} assesses the level of visual artifacts, particularly under low-light conditions where excessive noise may obscure important facial details.
    \item \textbf{Sharpness} assesses the clarity and definition of facial structures, which may be degraded by motion blur, defocus, or low resolution.
    \item \textbf{Colorfulness} assesses the accuracy of color representation, which may be affected by infrared interference, poor lighting, or colored light sources such as traffic signals or neon lights, leading to unnatural hues or color shifts.
    \item \textbf{Contrast} assesses the dynamic range between light and dark regions, with lighting variations often causing overexposed or underexposed areas that reduce overall contrast.
    \item \textbf{Fidelity} assesses the authenticity of facial content by identifying pseudo-structures or hallucinated details introduced by enhancement algorithms, which may compromise the realism and trustworthiness of the image.
    \item \textbf{Overall quality} assesses the general perceptual quality by integrating all the above dimensions, balancing distortion severity and the fidelity of facial content.
\end{itemize}

\begin{figure*}[!t]
 \setlength{\abovecaptionskip}{-5pt}

  \centering
  \includegraphics[width=0.92\textwidth]{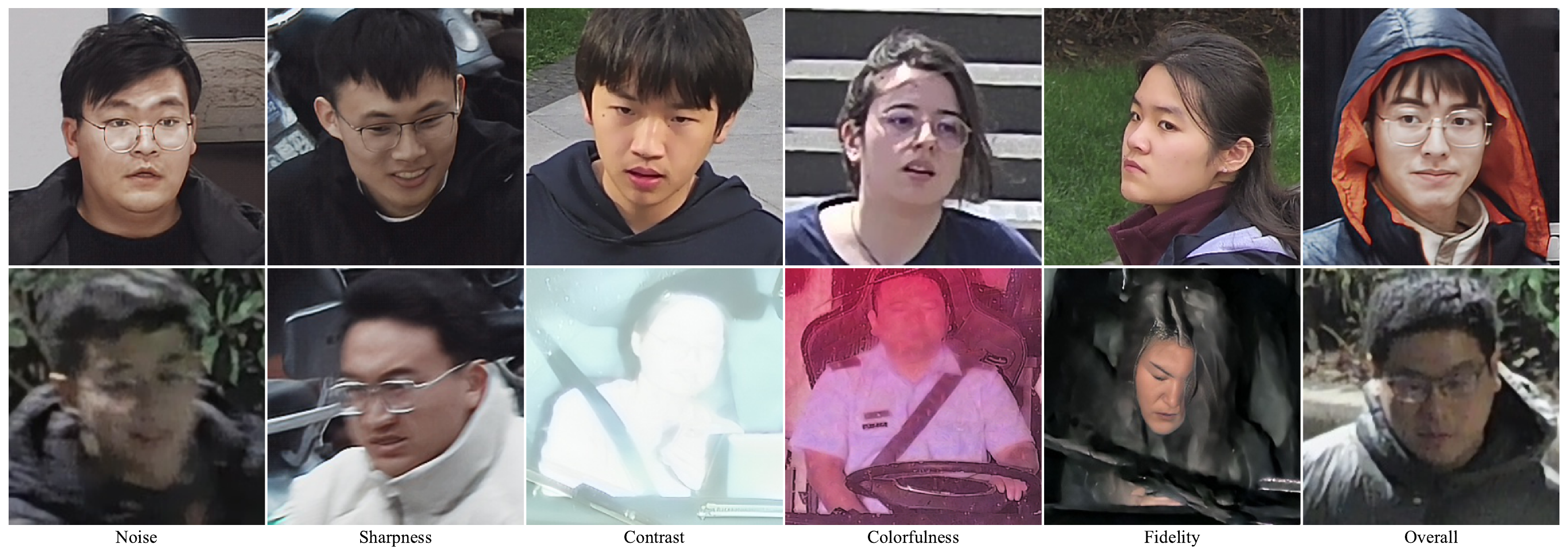}
  \caption{Representative surveillance facial images illustrating each quality dimension in SFIQA-Bench. The top row shows high-quality examples, while the bottom row presents low-quality counterparts for Noise, Sharpness, Contrast, Colorfulness, Fidelity, and Overall Quality, respectively.}
  \label{fig:dist}

\end{figure*}

To better understand these quality dimensions, we present several representative surveillance facial images in Fig.~\ref{fig:dist}, each exhibiting either high or low quality with respect to the corresponding dimension.

\subsubsection{Subjective Experiment Methodology}
We adopted the five-point Absolute Category Rating (ACR) scale for quality score annotation, following the standards in~\cite{hosu2020koniq, itu1999subjective}. The five categories---bad, poor, fair, good, and excellent---correspond to scores from 1 to 5, respectively. The subjective experiment for SFIQA-Bench was conducted in two phases: a \textbf{pilot study} and a \textbf{formal study}. In the pilot study, participants were required to complete a training session. They first watched a pre-recorded instructional video explaining common distortions in SFIs and the criteria for evaluating various quality dimensions. Afterwards, they rated 20 training samples that had been annotated by experts. Specifically, the expert team consists of one professor, one postdoctoral researcher specializing in IQA, two PhD candidates with research experience in IQA/FIQA, and two R\&D engineers from leading surveillance device manufacturers. All team members have either published peer-reviewed papers in the areas of IQA/FIQA or have participated in industry or academic projects related to surveillance imaging systems and quality evaluation. The expert scores were obtained through a collaborative expert annotation process. Participants were shown one image at a time, and after carefully observing it, they provided 6 scores before proceeding to the next image. If a participant’s Spearman Rank Correlation Coefficient (SRCC) with the expert scores for any quality dimension fell below 0.9, they were required to repeat the training until this threshold was met.

During the formal evaluation phase, we divide all test images into 25 sessions, each session consisted of 200 test images (except for one session, which contained 204). To ensure reliability, five ``golden" images---pre-annotated with either the best (score 5) or worst (score 1) quality---were included, along with five repeated images for rating consistency checking. In total, participants rated 208 images per session. If a participant's rating deviated by more than 1 point from the reference scores for the golden images, or if their ratings for the repeated images differed by more than 1 point, those scores were considered outliers. If the proportion of outlier scores exceeded 10\%, the participant's data for that session was discarded.

We recruited 100 participants for the subjective experiment, with each session rated by 25 participants, resulting in a total of 750,600 subjective ratings. Based on the golden image and repeated image checks, we excluded the ratings from 1 participants and 3 sessions due to inconsistency. The subjective scores were then processed in accordance with the ITU-T BT.500-13 recommendations \cite{bt2002methodology}, which include an outlier detection strategy based on the standard deviation of ratings and a standardized procedure for computing Mean Opinion Scores (MOS). Specifically, individual ratings that deviate significantly from the mean were identified as outliers and removed, and participants with more than 5\% outlier ratings were excluded from further analysis. In our experiment, a total of 894 ratings were removed(accounting for 0.125\% of the total ratings), and no participants were excluded. The original MOS computation defined by the ITU standard involves normalizing individual scores into Z-scores, linearly rescaling them to the [0, 100] range, and averaging across valid participants. In our case, we omit the Z-score rescaling to better align the final scores with the five predefined quality categories. Following the above procedures, we obtain the ground-truth quality scores for each dimension of every SFI in SFIQA-Bench.

\subsection{MOS Analysis}
\label{dataana}
We analyze the MOSs of SFIs in SFIQA-Bench from three apsects:

\subsubsection{\textbf{MOS Distributions}} 
\begin{figure*}[!t]
  \centering
  \includegraphics[width=0.95\linewidth]{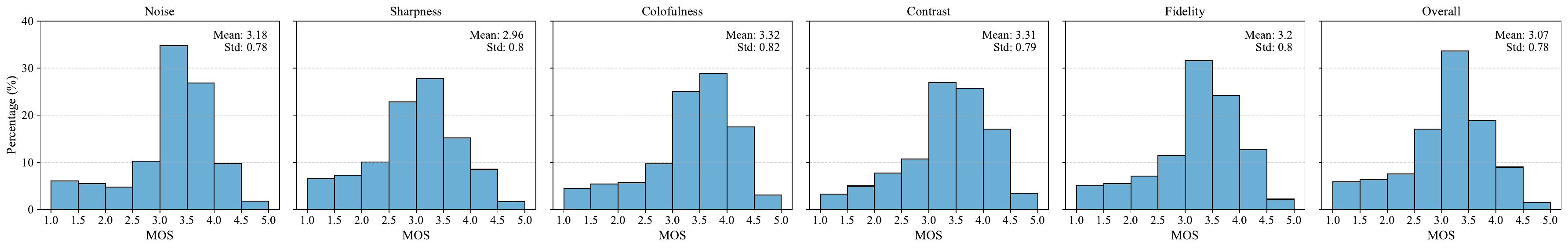}
  \caption{MOS distributions for the six quality dimensions in SFIQA-Bench: Noise, Sharpness, Colorfulness, Contrast, Fidelity, and Overall Quality.}
  \label{fig:hist_6dims}
\end{figure*}

We illustrate the MOS distributions of the six quality dimensions in Fig.~\ref{fig:hist_6dims}. Across all dimensions, the MOS values exhibit approximately normal or slightly left-skewed distributions, with most ratings concentrated between 2.5 and 4.0. The mean MOS ranges from 2.96 (sharpness) to 3.32 (colorfulness), indicating that the quality of most SFIs is moderate, with noticeable variation across dimensions. 

Notably, sharpness has the lowest average score, suggesting that the loss of facial texture and structure remains a dominant distortion in SFIs. This is largely due to factors such as pedestrian motion, in-vehicle monitoring, and low resolution. In contrast, colorfulness and contrast achieve relatively higher MOS values, indicating that modern surveillance cameras tend to perform better in these aspects. Noise and fidelity exhibit similar distributions, likely because noise commonly occurs under low-light conditions, which in turn activates facial enhancement modules. These enhancement processes, particularly those based on deep learning, may introduce pseudo-structures that compromise facial fidelity.

Interestingly, the overall quality distribution is closer to a Gaussian distribution than those of the individual dimensions, while still exhibiting a slight left skew. This can be explained by the fact that overall quality integrates all five individual dimensions, leading to a natural averaging effect that smooths out skewness. Overall, the results highlight the diverse range of quality issues present in the benchmark and underscore the need for multidimensional quality assessment in real-world surveillance scenarios.

\subsubsection{\textbf{The Correlations of Different Dimensions}} 
\begin{figure}[!t]
  \centering
  \includegraphics[width=0.8\linewidth]{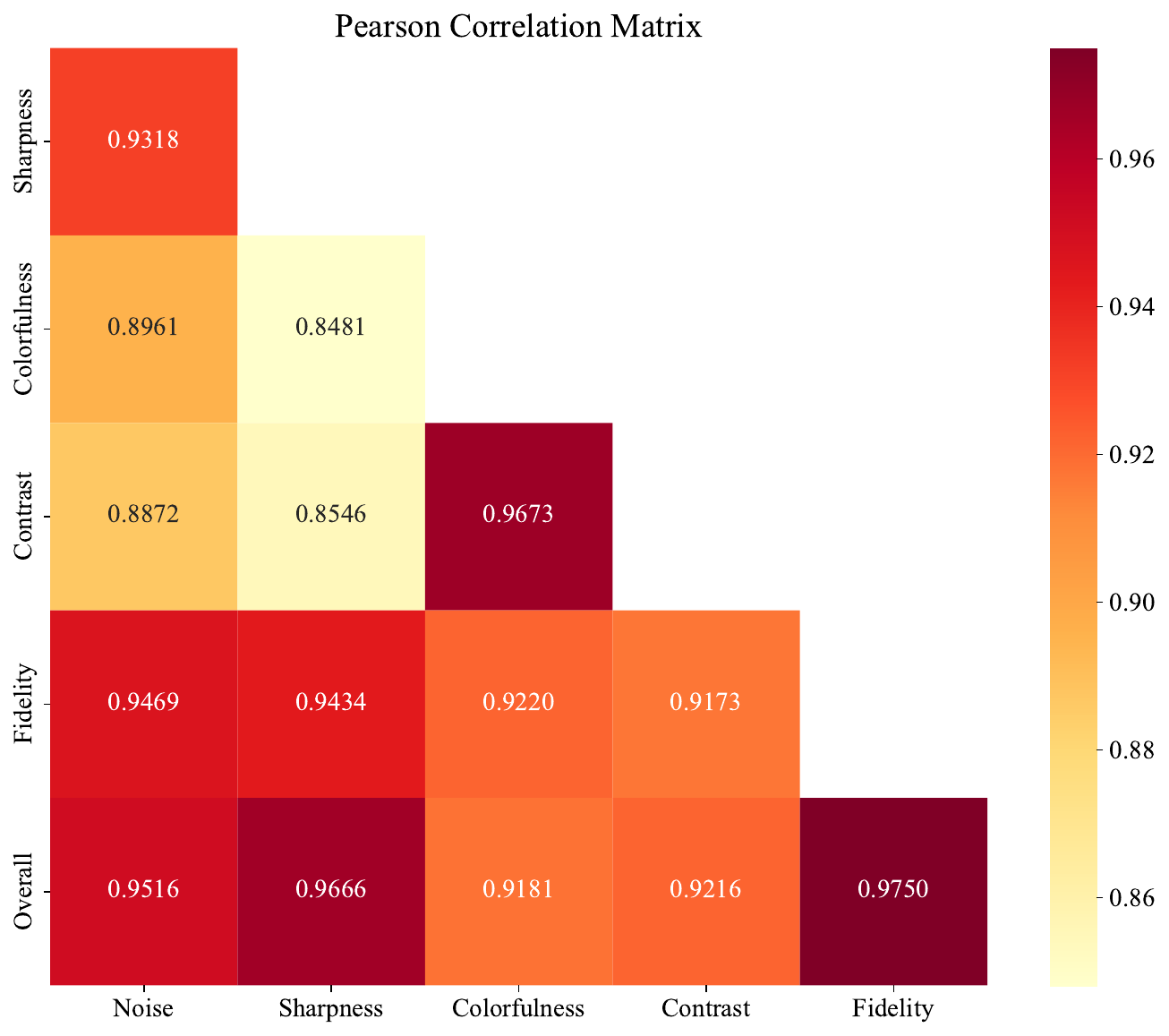}
  \caption{The matrix of PLCCs of different dimensions. Red stands for high PLCC and yellow stands for low.}
  \label{fig:plcc_doubledouble}
\end{figure}

Fig.~\ref{fig:plcc_doubledouble} presents the Pearson correlation coefficients (PLCC) among the six quality dimensions. Overall quality exhibits strong correlations with all five individual dimensions, with the highest correlation observed with fidelity, followed by sharpness and noise. This suggests that the authenticity of facial structure, image clarity, and noise level are the most influential factors in shaping subjective judgments of overall quality. 

Among the five perceptual dimensions, fidelity is strongly correlated with both noise and sharpness. The relationship between fidelity and noise has been previously discussed in the MOS distribution analysis. As for the link between fidelity and sharpness, this may be attributed to the fact that facial structures are composed of fine-grained textures, which inherently influence perceived sharpness. Contrast and colorfulness also exhibit a high correlation, suggesting that these attributes tend to vary together, likely due to shared dependencies on environmental illumination intensity and color rendering. In contrast, colorfulness and contrast show weaker correlations with noise and sharpness, which may be explained by their differing nature---color and contrast are more global perceptual attributes, while noise and sharpness relate to local structural distortions. 

Finally, we conduct a multiple linear regression analysis using the least squares method~\cite{bjorck2024numerical} to quantify the contribution of each quality dimension to the overall quality. The fitted model is expressed as:
\begin{align}
\text{Overall} ={} & 0.0765 \cdot \text{Noise} + 0.4134 \cdot \text{Sharpness} \notag \\
                   &+ 0.0554 \cdot \text{Colorfulness} + 0.1463 \cdot \text{Contrast} \notag \\
                   &  + 0.3076 \cdot \text{Fidelity}
\end{align}
This model yields a high coefficient of determination ($R^2=0.9778$), indicating that the five perceptual dimensions explain $97.78\%$ of the variance in overall quality scores. The regression results highlights the dominant contributions of sharpness and fidelity in determining overall quality, followed by contrast, noise, and colorfulness. These findings are general consistent with our PLCC analysis results.

\begin{figure*}[!t]
        \centering
        \includegraphics[width=0.92\textwidth]{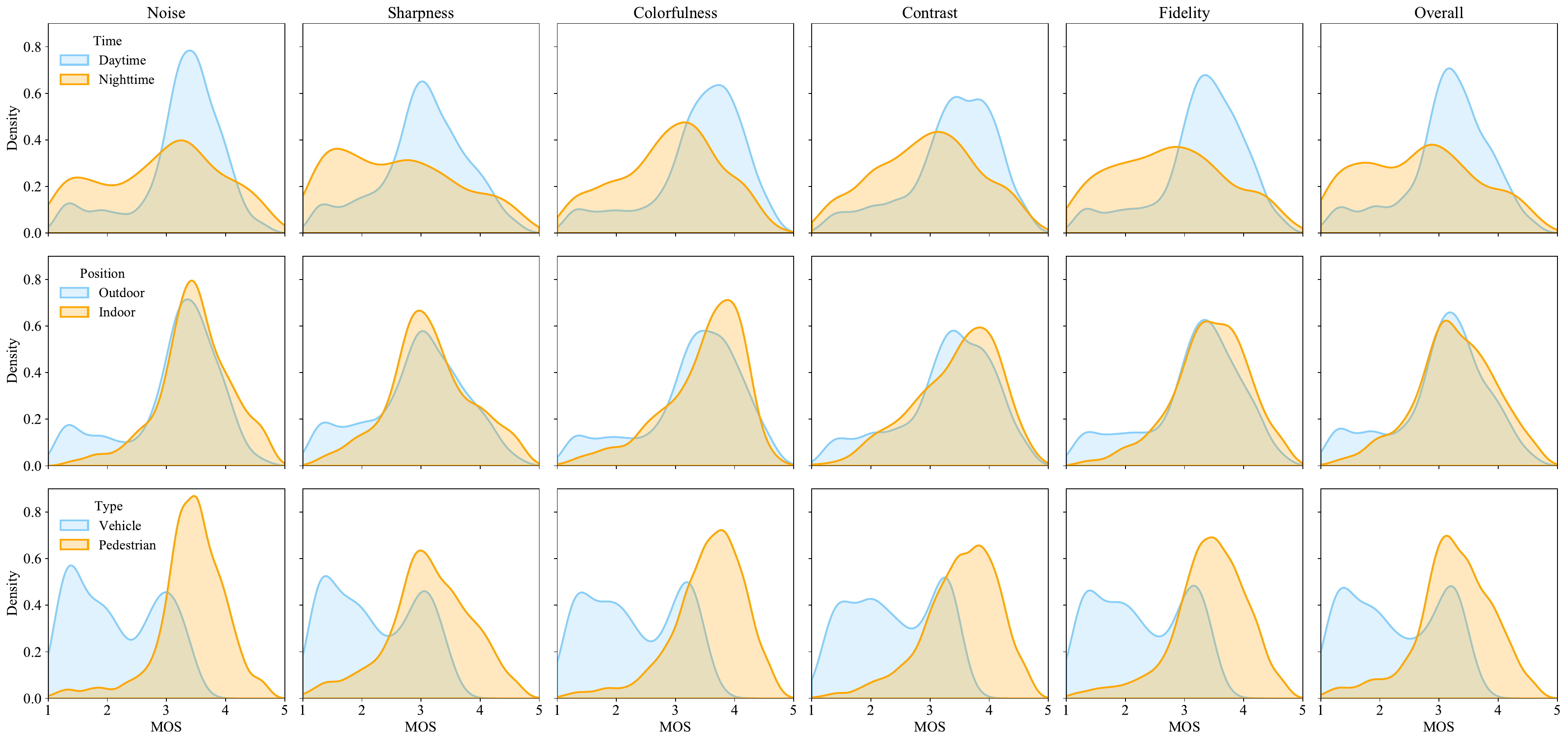}
        \caption{Probability density distributions of MOSs across different surveillance scenarios in SFIQA-Bench, including daytime vs. nighttime, indoor vs. outdoor, and vehicle vs. pedestrian conditions.}
        \label{fig:kde}
    \end{figure*}

\subsubsection{\textbf{MOS Distributions for Different Scenarios}} To analyze how environmental and capture conditions affect perceived quality, we visualize the probability density distributions of MOSs across different scenarios, including daytime vs. nighttime, pedestrian vs. vehicle, and indoor vs. outdoor conditions in Fig.~\ref{fig:kde}. 

For daytime vs. nighttime, images captured during the daytime generally exhibit higher MOS across all quality dimensions. In particular, daytime images show a clear shift toward higher scores in sharpness, fidelity, contrast, and overall quality, reflecting the benefits of adequate ambient lighting. Conversely, nighttime images suffer from increased noise and reduced sharpness, likely due to low illumination and camera gain adjustments, which degrade the quality of facial structure representation. 

For vehicle vs. pedestrian, images captured in vehicle-based surveillance scenarios (e.g., driver monitoring) tend to have lower quality than those captured in pedestrian contexts. The probability density distributions figures show vehicle images skewing toward lower scores across noise, sharpness, fidelity, and overall quality. This is attributed to challenging conditions such as motion blur, window reflections, uneven lighting, and oblique camera angles common in in-car environments. In contrast, pedestrian images---often captured in more open and stable conditions---achieve relatively higher quality scores. 

For indoor vs. outdoor, indoor scenes typically yield better quality scores than outdoor ones, especially for fidelity, sharpness, and contrast. The more stable and controlled lighting indoors leads to consistent image exposure and fewer distortions. Outdoor scenes, by comparison, are more prone to varying light intensities, backlighting, and environmental interferences, which result in broader distributions and lower average MOS in several dimensions.

\begin{figure*}[!t]

  \centering
  \includegraphics[width=\textwidth]{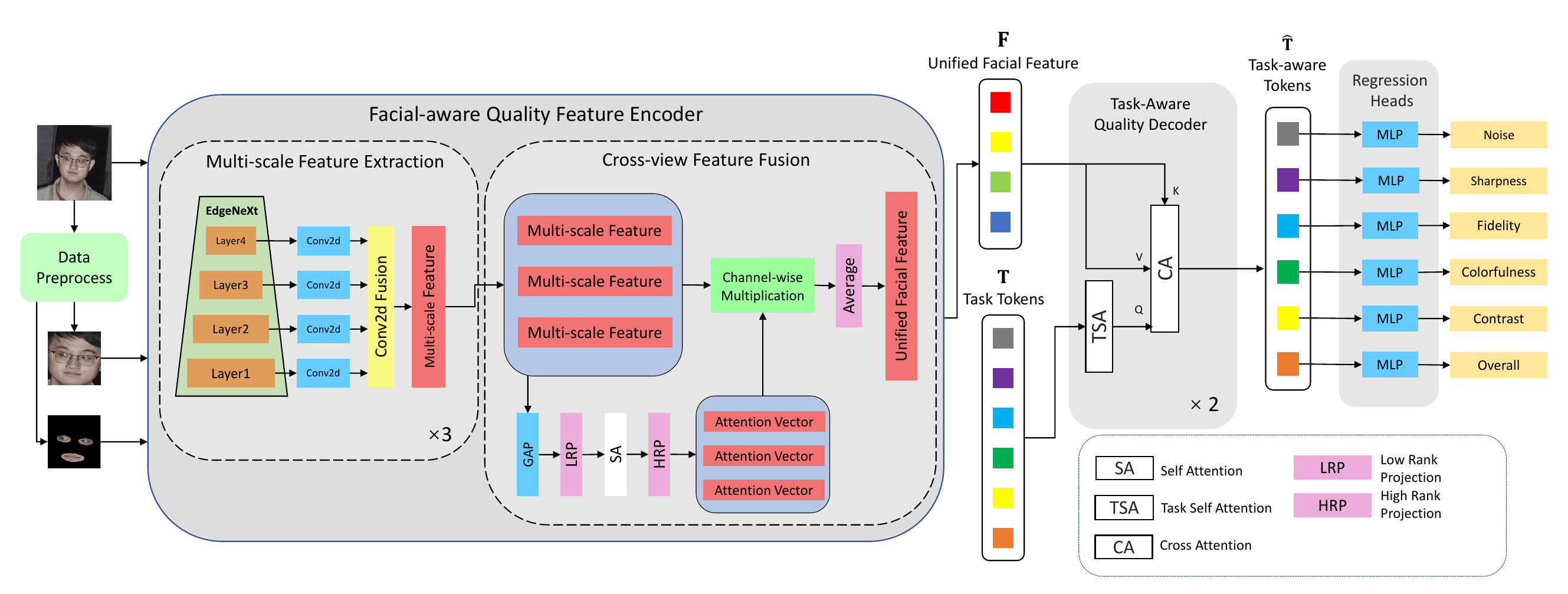}
  \caption{Overview of our proposed framework. The method contains a preprocess module, a facial-aware quality feature encoder, a task-aware decoder and prediction heads. Input image are preprocessed to compact face and eyes-and-mouth region. The facial-aware quality feature encoder extracts multi-scale, inter-view facial quality features $\mathbf{F}$. Task tokens $\mathbf{T}$ are fused with unified facial features $\mathbf{F}$ in the decoder via task-specific and cross-attention mechanisms. The resulting task-aware tokens are then passed through regression heads to predict six quality scores: noise, sharpness, fidelity, colorfulness, contrast, and overall quality.}
  \label{fig:method}
\end{figure*}

\section{SFIQA-Assessor}
In this section, we present SFIQA-Assessor, a multi-dimensional objective quality assessment model for SFIs. Considering that SFIQA models are often deployed on resource-constrained client-side devices, such as surveillance cameras, the model is carefully optimized to achieve high accuracy while maintaining low computational complexity, making it well-suited for real-world deployment.

\subsection{Data Preprocessing}
\begin{figure}[!t]
\setlength{\abovecaptionskip}{-10pt}
  \centering
  \includegraphics[width=\linewidth]{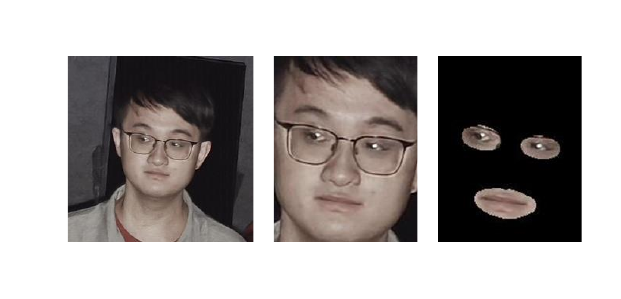}
  \caption{Preprocessed images: the original loosely cropped image, the compact face region, the eyes-and-mouth region.}
  \label{fig:threeface}
\end{figure}

Previous studies have shown that, in the perceptual quality assessment of facial images, the facial region is more important than non-facial areas \cite{liu2024assessing}. Among various facial components, the eyes and mouth have the greatest influence on perceived facial quality \cite{faure2002influence, liu2024assessing}. To explicitly leverage this critical information, we perform a preprocessing procedure to extract both a compact face region and an eyes-and-mouth region. Specifically, given an original SFI $\bm{x}_o$, automatically captured by surveillance cameras, we first employ the Ultra-Light-Fast-Generic-Face-Detector-1MB tool~\cite{ultraface1mb} to detect a tightly cropped face bounding box and obtain the compact face region $\bm{x}_{f}$. Next, we apply a face parsing algorithm~\cite{narayan2025segface} to $\bm{x}_{f}$ to segment facial components and retain only the eyes and mouth regions, resulting in $\bm{x}_{e,m}$. Finally, all three facial views---the original loosely cropped image $\bm{x}_o$, the compact face region $\bm{x}_f$, and the eyes-and-mouth region $\bm{x}_{e,m}$---are fed into our model, enabling it to incorporate both contextual background information and the most perceptually salient facial features. An example of these three inputs is shown in Fig.~\ref{fig:threeface}.

\subsection{Facial-aware Quality Feature Encoder}
\label{Sec:SCFE}

After obtaining three facial views, we design a facial-aware quality feature encoder to extract quality-relevant features with a focus on facial regions, enabling effective multi-dimensional quality assessment.

\subsubsection{Multi-scale Feature Extraction} 
We adopt a lightweight CNN model, EdgeNeXt~\cite{maaz2022edgenext}, as the default backbone to extract rich visual feature maps (other backbones are discussed in Section~\ref{sec:abl}). Prior studies~\cite{zeiler2014visualizing} have shown that shallow layers of CNNs respond primarily to low-level visual features, while deeper layers capture high-level semantic information. In the context of our multi-dimensional quality assessment task, dimensions such as contrast, noise, colorfulness, and sharpness are closely related to low-level features~\cite{mittal2012no}, whereas fidelity relies more on high-level semantic representations~\cite{zhang2018unreasonable}. Therefore, we employ a hierarchical multi-scale feature extraction strategy that aggregates feature maps from all intermediate layers of EdgeNeXt to effectively support multi-dimensional quality prediction.

Specifically, for three facial views $\bm{x}_o$, $\bm{x}_f$, and $\bm{x}_{e,m}$, we feed each into the backbone model $\mathcal{B}$ to obtain multi-scale feature representations. Let $N_s$ denote the number of stages in $\mathcal{B}$. The feature maps at scale $i$ are given by:

\begin{align}
\mathbf{F}^{(i)}_j = \mathcal{B}^{(i)}(\bm{x}_j), 
\end{align}
where $\mathbf{F}^{(i)}_j$ denotes the feature maps from the $i$-th stage of the backbone for view $j$, $i \in \{1, 2, \dots, N_s\}$ indexes the stages of the backbone network, and $j \in \{o,f,(e,m)\}$ corresponds to three facial views.

For the multi-scale features extracted from each view, we first apply a multi-layer perceptron (MLP) to project the feature maps from different scales to a unified channel dimension:
\begin{equation}
   \mathbf{\hat{F}}^{(i)}_j = \mathrm{MLP_{D_i\rightarrow D_o}}(\mathbf{F}^{(i)}_j),
\end{equation}
where $D_i$ and $D_o$ represent the input and output channel dimensions, respectively. The projected feature maps across all scales are then concatenated along the channel dimension and fused using a $1\times1$ CNN:
\begin{equation}
    \mathbf{F}_j = \mathrm{Conv}^{1 \times 1}_\mathrm{(N_s*D_o\rightarrow D_o)} (\mathrm{Concat}(\mathbf{\hat{F}}^{(1)}_j,\mathbf{\hat{F}}^{(2)}_j,...,\mathbf{\hat{F}}^{(N_s)}_j)),
\end{equation}
where $\mathbf{F}_j$ denotes the final fused multi-scale feature representation for view $j$.

\subsubsection{Cross-view Feature Fusion}
\label{sec:LCRA}

We then fuse the extracted multi-scale features from all views to obtain a unified facial-aware quality feature representation. To enable effective cross-view interactions, we employ an attention-based fusion strategy to integrate features across different views. However, standard attention mechanisms suffer from a quadratic computational complexity of $O(n^2)$, which contradicts our lightweight design principle. To address this, we adopt a \textbf{low-rank channel attention} approach. Specifically, each feature map is first compressed into a global semantic vector along the spatial dimensions via global average pooling:
\begin{equation}
    \mathbf{\mathit{f}}_j 
= \mathrm{GAP}(\mathbf{F}_j)
= \frac{1}{H\,W}\sum_{h=1}^{H}\sum_{w=1}^{W} \mathbf{F}_j[:,h,w]
\;\in\mathbb{R}^{D_o},
\end{equation}
where $H$ and $W$ denote the height and width of the feature map, respectively. This vector is then projected into a lower-dimensional embedding space using a learnable linear transformation. This process is referred to as Low-Rank Projection (LRP):
\begin{equation}
    \widetilde{\mathbf{\mathit{f}}}_j 
= \mathbf{W}_d \mathbf{\mathit{f}}_i  
\in\mathbb{R}^{D_l}.
\end{equation}
where $\textbf{W}_d \in \mathbb{R}^{D_o\times D_l}$ is a learnable projection matrix, and $D_l$ is the latent dimension of each projected feature vector. The projected vectors $\{\widetilde{\mathbf{\mathit{f}}}_i\}$ from all views are concatenated along the sequence (view) dimension to form the input embedding for the attention module:
\begin{equation}
\widetilde{\mathbf{F}} = [\widetilde{\mathbf{\mathit{f}}}_o; \widetilde{\mathbf{\mathit{f}}}_f; \widetilde{\mathbf{\mathit{f}}}_{e,m}] \in \mathbb{R}^{3\times D_l}.
\end{equation}

Next, we introduce a self-attention module to model the inter-dependencies among the global vectors $\widetilde{\mathbf{F}}$ obtained from the three facial views. By applying attention over this sequence, the network learns to capture complementary information across different views. The self-attention mechanism is defined as follows:
\begin{align}
\mathbf{G} &= \mathrm{Attn}(\mathbf{Q}, \mathbf{K}, \mathbf{V})= \mathrm{softmax}\!\left( \frac{QK^\top}{\sqrt{d_h}} \right) V, \nonumber \\
\mathbf{Q} &= \mathbf{K} = \mathbf{V} = \widetilde{\mathbf{F}},\quad \mathbf{G} \in \mathbb{R}^{ 3 \times D_l},
\end{align}
where $\mathbf{G}$ represents the attention–fused feature sequences. To enable cross-view feature fusion, $\mathbf{G}$ must be up-projected to match the original number of channels. This up-projection process can be formulated as follows, which is referred to as High-Rank Projection (HRP):
\begin{equation}
    \mathbf{U}
= \mathbf{W}_u\mathbf{G}\in\mathbb{R}^{3 \times D_o},
\end{equation}
where $\mathbf{W}_u\in\mathbb{R}^{D_l\times D_o}$ is a lenrnable preojection matrix. 

Finally, the cross-view fusion operation is performed to obtain the unified facial feature representation $\mathbf{F}$, computed as:
\begin{equation}
    \mathbf{F} = \frac{1}{3} \sum_{j=1}^3 \mathbf{F}_j \odot \mathbf{\textit{u}}_j \in \mathbb{R}^{ D_o \times H \times W},
\end{equation}
where $\mathbf{\textit{u}}_j$ denotes the $j$-th attention vector from $\mathbf{U}$, and $\odot$ represents channel-wise multiplication broadcast across the spatial dimensions.

\subsection{Task-Aware Quality Decoder}

Since our model tackles a multi-task learning problem involving the regression of quality scores across multiple dimensions, we design a task-aware quality decoder to disentangle the unified facial feature representation $\mathbf{F}$ into task-specific feature vectors corresponding to each quality dimension. Specifically, we introduce a set of learnable task tokens $\mathbf{T}$, which serve as queries in a cross-attention module to interact with the shared facial feature representation $\mathbf{F}$. As discussed in Section~\ref{dataana}, the tasks associated with different quality dimensions are not entirely independent but exhibit inherent correlations. To capture these relationships, the task tokens $\mathbf{T}$ are first processed through a task self-attention module, allowing information exchange across tasks. The updated task tokens $\mathbf{T'}$ are computed as:
\begin{equation}
    \mathbf{T'} = \mathrm{Attn}(\mathbf{Q} = \mathbf{T},\mathbf{K} = \mathbf{T},\mathbf{V} = \mathbf{T}).
\end{equation} 

The updated tokens $\mathbf{T'}$ are then used as queries in the cross-attention module to attend to the facial feature representation $\mathbf{F}$. The output of the cross-attention module, denoted as task-aware tokens $\mathbf{\hat{T}}$, is computed as: 
\begin{equation}
    \mathbf{\hat{T}} = \mathrm{Attn}(\mathbf{Q} = \mathbf{T'},\mathbf{K} = \mathbf{F},\mathbf{V} = \mathbf{F}).
\end{equation}

To enhance performance, the decoder, comprising both the task self-attention and cross-attention modules, is applied twice. In the first pass, task tokens interact with the shared facial features to obtain coarse task-specific cues. The second pass further refines these tokens by re-modeling inter-task relationships and re-attending to the shared features, leading to more stable and discriminative representations for each quality dimension. After this two-stage decoding process, the final task-aware tokens $\mathbf{\hat{T}}$ are obtained. These tokens not only preserve the task-invariant representations learned through the task self-attention mechanism, but also capture task-specific facial features through the cross-attention interaction.

\subsection{Regression Heads}
We employ a three-layer MLP with 128, 128, and 1 neurons as the regression head to map each task-specific token to its corresponding quality score dimension:
\begin{equation}
\hat{q}_k = \mathrm{MLP}(\mathbf{\hat{\mathit{t}}}_k), \quad k \in [1,2,..,6],
\end{equation}
where $\mathbf{\hat{\mathit{t}}}_k$ denotes the $k$-th task token from $\mathbf{\hat{T}}$, and $\hat{q}_k$ is the output quality score for the $k$-th dimension of the input facial images.

Regarding the loss function, we assume equal importance across all tasks. Therefore, we compute the mean squared error (MSE) between the predicted scores and ground truth labels for each quality dimension, and average them to obtain the final loss:
\begin{equation}
\mathcal{L} = \frac{1}{6}\sum^6_{k=1} (q_k - \hat{q}_k)^2,
\end{equation}
where $\mathcal{L}$ denotes the total loss for a facial image, and $q_k$ represents the MOS for the $k$-th quality dimension.

\section{Experiments}
\subsection{Validation Datasets}
We first evaluate our SFIQA-Assessor on SFIQA-Bench. To further assess its generalization capability, we conduct experiments on two widely used perceptual FIQA datasets: GFIQA-20K~\cite{su2023going} and CGFIQA-40K~\cite{chen2024dsl}.  A detailed introduction of these datasets is provided in Table~\ref{tab:Datasets}. For SFIQA-Bench, we perform five‑fold cross‑validation, splitting each fold into training, validation, and test sets in a 7:1:2 ratio. For GFIQA-20K, we follow the data split protocol used in DSL-FIQA~\cite{chen2024dsl}, adopting a 7:1:2 ratio for training, validation and testing. For CGFIQA-40K, we use the official training, validation, and test split.

\begin{table*}[!t]
\centering
\setlength{\tabcolsep}{3pt}
\renewcommand{\arraystretch}{1}
\caption{Comparison of Different IQA Methods on Proposed SFIQA-Bench Dataset. Bold indicates the 1st and underline the 2nd. Methods highlighted in yellow represent generic IQA approaches, those in gray denote recognition-oriented FIQA methods, and those in blue indicate perceptual FIQA methods.}
\label{tab:method_comparison}
\resizebox{\textwidth}{!}{
\begin{tabular}{c*{12}{c}ccc}
\toprule
\multirow{2}{*}{\textbf{Method}}
  & \multicolumn{2}{c}{\textbf{Noise}}
  & \multicolumn{2}{c}{\textbf{Sharpness}}
  & \multicolumn{2}{c}{\textbf{Colorfulness}}
  & \multicolumn{2}{c}{\textbf{Contrast}}
  & \multicolumn{2}{c}{\textbf{Fidelity}}
  & \multicolumn{2}{c}{\textbf{Overall}}
  & \multirow{2}{*}{\textbf{\makecell{Params\\(M)}}}
  & \multirow{2}{*}{\textbf{\makecell{GMACs}}}
  & \multirow{2}{*}{\textbf{\makecell{Latency\\(ms)}}} \\
\cmidrule(lr){2-3}
\cmidrule(lr){4-5}
\cmidrule(lr){6-7}
\cmidrule(lr){8-9}
\cmidrule(lr){10-11}
\cmidrule(lr){12-13}
 & SRCC & PLCC & SRCC & PLCC & SRCC & PLCC & SRCC & PLCC & SRCC & PLCC & SRCC & PLCC
 & & & \\
\midrule
\rowcolor{yellow!25}BRISQUE\cite{mittal2012no}  & 0.7553 & 0.8442 & 0.7506 & 0.7813 & 0.7041 & 0.7942 & 0.7075 & 0.7699 & 0.7147 & 0.7913 & 0.7330 & 0.7895 & - & - & - \\
\rowcolor{yellow!25}NIQE\cite{mittal2012making} & 0.0242 & 0.1547 & 0.0067 & 0.0704 & 0.0168 & 0.1273 & 0.0192 & 0.1115 & 0.0170 & 0.1118 & 0.0150 & 0.1020 & - & - & - \\
\rowcolor{yellow!25}HyperIQA\cite{su2020blindly}
  & 0.7765 & 0.8556 & 0.8011 & 0.8372 & 0.7831 & 0.8553 & 0.7903 & 0.8497 & 0.7522 & 0.8460 & 0.7679 & 0.8434 & 27.38  & 107.83  & 26.41  \\
\rowcolor{yellow!25}TReS\cite{golestaneh2021no}
  & 0.8482 & 0.8919 & 0.8680 & 0.8805 & 0.8411 & 0.8794 & 0.8486 & 0.8699 & 0.8374 & 0.8801 & 0.8481 & 0.8796 & 152.45 & 500.04  & 158.67 \\
\rowcolor{yellow!25}MUSIQ\cite{ke2021musiq}
  & 0.8828 & 0.9114 & 0.9076 & 0.9200 & 0.7638 & 0.7719 & 0.8487 & 0.8830 & 0.8731 & 0.8975 & 0.9035 & 0.9201 & 78.59  & 115.37  & 72.09  \\
\rowcolor{yellow!25}MANIQA\cite{yang2022maniqa}
  & 0.8058 & 0.8947 & 0.8451 & 0.8766 & 0.8620 & 0.9137 & 0.8420 & 0.8872 & 0.7937 & 0.8591 & 0.8491 & 0.9029 & 135.75 & 2580.2  & 402.03 \\
\rowcolor{yellow!25}StairIQA\cite{sun2023blind}
  & 0.9224 & 0.9437 & 0.9335 & 0.9426 & 0.9174 & 0.9391 & 0.9237 & 0.9431 & 0.9168 & 0.9442 & 0.9286 & 0.9472 & 30.49  & 10.43 & 31.02 \\
\rowcolor{yellow!25}CLIPIQA\cite{wang2023exploring}
  & 0.6591 & 0.6828 & 0.6401 & 0.5817 & 0.6828 & 0.7031 & 0.6940 & 0.7093 & 0.6432 & 0.5392 & 0.6413 & 0.5910 & 152.45 & 35.25   & 20.82  \\
\rowcolor{yellow!25}LIQE\cite{zhang2023blind}
  & 0.9162 & 0.9430 & 0.9331 & 0.9447 & 0.9270 & 0.9473 & \underline{0.9320} & 0.9502 & 0.9148 & 0.9385 & 0.9290 & 0.9463 & 151.28 & 1506.88 & 296.10 \\
\rowcolor{yellow!25}TOPIQ\cite{chen2024topiq}
  & 0.8874 & 0.9168 & 0.8870 & 0.9050 & 0.8787 & 0.9302 & 0.8927 & 0.9226 & 0.8481 & 0.8893 & 0.8768 & 0.9148 & 45.2   & 20.88   & 22.74  \\
\rowcolor{gray!25}SER-FIQ\cite{terhorst2020ser}
  & 0.5058 & 0.5445 & 0.4408 & 0.4748 & 0.5258 & 0.5505 & 0.4985 & 0.5304 & 0.4912 & 0.5189 & 0.4697 & 0.5055 & 43.72  & 6.35  & 29.15 \\
\rowcolor{gray!25}SDD-FIQA\cite{ou2021sdd}
  & 0.5872 & 0.6109 & 0.5996 & 0.6369 & 0.5891 & 0.6206 & 0.6102 & 0.6358 & 0.6218 & 0.6473 & 0.5905 & 0.6259 & 30.78  &  6.31 & 24.87 \\
\rowcolor{lightblue!20} IFQA\cite{jo2023ifqa}
&0.8626 &	0.8869 &	0.8849 &	0.9017 &	0.8776 &	0.9012 &	0.8724 	&0.8961 	&0.8812 	&0.9079 	&0.8928 &	0.9273 	&105.17	&26.58	&28.37 \\
\rowcolor{lightblue!20}DN-PIQA\cite{sun2024dual}
  & 0.8424 & 0.8895 & 0.8745 & 0.8885 & 0.8750 & 0.9088 & 0.9059 & 0.9208 & 0.8514 & 0.8915 & 0.8846 & 0.9071 & 173.69 & 89.32 & 80.28 \\
\rowcolor{lightblue!20}DSL-FIQA\cite{chen2024dsl}
  & \textbf{0.9311} & \textbf{0.9532} & \textbf{0.9422} & \textbf{0.9519}
  & \underline{0.9360} & \underline{0.9517} & \textbf{0.9417} & \textbf{0.9557} & \textbf{0.9274} & \textbf{0.9472} & \textbf{0.9411} & \textbf{0.9585} & 252.69 & 2427.2  & 518.90 \\
  \hdashline
\rowcolor{lightblue!20}SFIQA-Assessor-XXS & 0.9142 & 0.9453 & 0.9294 & 0.9417 & 0.9319 & 0.9438 & 0.9200 & 0.9488 & 0.9160 & 0.9473 & 0.9310 & 0.9499 & \textbf{5.28} & \textbf{1.39} & \textbf{8.26} \\
\rowcolor{lightblue!20}SFIQA-Assessor-XS  & 0.9178 & 0.9480 & 0.9343 & 0.9454 & 0.9324 & 0.9446 & 0.9192 & 0.9488 & 0.9177 & 0.9471 & 0.9344 & 0.9519 & \underline{8.26} & \underline{2.05} & \underline{9.34} \\
\rowcolor{lightblue!20}SFIQA-Assessor-S
  & \underline{0.9254} & \underline{0.9522} & \underline{0.9386} & \underline{0.9485} & \textbf{0.9376 }& \textbf{0.9525} & 0.9290 & \underline{0.9522} & \underline{0.9218} & \underline{0.9465} & \underline{0.9379} & \underline{0.9538} & 17.76  & 3.75   & 10.59 \\

\bottomrule
\end{tabular}
}
\end{table*}

\subsection{Compared Methods}
We compare the proposed SFIQA-Assessor with two categories of IQA methods: general-purpose IQA methods and face-specific IQA methods. The general-purpose methods include BRISQUE~\cite{mittal2012no}, NIQE~\cite{mittal2012making},  StairIQA~\cite{sun2023blind}, HyperIQA~\cite{su2020blindly}, MANIQA~\cite{yang2022maniqa}, MUSIQ~\cite{ke2021musiq}, CLIP-IQA~\cite{wang2023exploring}, TOPIQ~\cite{chen2024topiq}, LIQE~\cite{zhang2023blind}, and TReS~\cite{golestaneh2021no}. The face-specific methods include SER-FIQ~\cite{terhorst2020ser}, SDD-FIQA~\cite{ou2021sdd}, IFQA~\cite{jo2023ifqa}, DN-PIQA~\cite{sun2024dual}, and DSL-FIQA~\cite{chen2024dsl}. Among these, SER-FIQ and SDD-FIQA are recognition-oriented FIQA methods, whereas IFQA, DN-PIQA and DSL-FIQA are perceptual FIQA methods. For fairness, all methods except NIQE and SER-FIQ~\footnote{NIQE and SER-FIQ are opinion-unaware (unsupervised) method that does not require training.} are retrained on the three test datasets.

\begin{table}[!t]
\centering
\caption{Table of Performance on GFIQA-20K and CGFIQA-40K. }
\label{tab:GFIQA}
\begin{tabular}{ccccc}

\toprule
\multirow{2}{*}{\textbf{Method}} & \multicolumn{2}{c}{\textbf{GFIQA-20K}} & \multicolumn{2}{c}{\textbf{CGFIQA-40K}} \\
\cmidrule(lr){2-3} \cmidrule(lr){4-5}
               & SRCC & PLCC & SRCC & PLCC \\
\midrule
\rowcolor{yellow!25}HyperIQA\cite{su2020blindly}    & 0.8923 & 0.8846 & 0.9766 & 0.9629 \\
\rowcolor{yellow!25}TReS\cite{golestaneh2021no}     & 0.7967 & 0.7670 & 0.8505 & 0.8452 \\
\rowcolor{yellow!25}MUSIQ\cite{ke2021musiq}         & 0.9243 & 0.9203 & 0.9694 & 0.9701 \\
\rowcolor{yellow!25}MANIQA\cite{yang2022maniqa}     & 0.9639 & 0.9589 & 0.9521 & 0.9561 \\
\rowcolor{yellow!25}CLIPIQA\cite{wang2023exploring} & 0.8484 & 0.8519 & 0.8062 & 0.8109 \\
\rowcolor{yellow!25}StairIQA\cite{sun2023blind}     & 0.9624 & 0.9665 & 0.9831 & 0.9843 \\
\rowcolor{yellow!25}LIQE\cite{zhang2023blind}       & 0.9622 & 0.9405 & 0.9839 & 0.9567 \\
\rowcolor{yellow!25}TOPIQ\cite{chen2024topiq}       & 0.9641 & 0.9581 & \underline{0.9844} & 0.9806 \\
\rowcolor{gray!25}SER-FIQ\cite{terhorst2020ser}  & 0.4873 & 0.5239 & 0.6037 & 0.6272 \\
\rowcolor{gray!25}SDD-FIQA\cite{ou2021sdd}        & 0.5633 & 0.5938 & 0.7030 & 0.7257 \\
\rowcolor{lightblue!20} IFQA\cite{jo2023ifqa}   & 0.9514	&0.9533	&0.9779	&0.9758 \\
\rowcolor{lightblue!20}DN-PIQA\cite{sun2024dual}       & 0.9430 & 0.9379 & 0.9714 & 0.9724 \\
\rowcolor{lightblue!20}DSL-FIQA\cite{chen2024dsl}      & \textbf{0.9696} & \textbf{0.9699} & \textbf{0.9861} & \textbf{0.9865} \\
\rowcolor{lightblue!20}SFIQA-Assessor-S                        & \underline{0.9654} & \underline{0.9666} & 0.9842 & \underline{0.9846} \\

\bottomrule
\end{tabular}
\end{table}

\subsection{Training Settings}
The implementation of SFIQA-Assessor is based on PyTorch. We adopt EdgeNeXt~\cite{maaz2022edgenext} as the lightweight backbone and explore three variants: EdgeNeXt-S, EdgeNeXt-XS, and EdgeNeXt-XXS, denoted as SFIQA-Assessor-S, SFIQA-Assessor-XS, and SFIQA-Assessor-XXS, respectively. Unless otherwise specified, SFIQA-Assessor-S is used as the default configuration. The backbone consists of $N_s = 4$ stages for EdgeNeXt, and $N_s$ follows the stage configuration of the selected backbone. The channel dimensions for the output and latent features, $D_o$ and $D_l$, are set to 128 and 64, respectively. For the self-attention modules, the embedding dimension is $64$ with $4$ attention heads. The task-specific self-attention and cross-attention modules use an embedding dimension of $128$ and $8$ attention heads. Input facial images are resized to $224 \times 224$. The pretrained weights on ImageNet~\cite{deng2009imagenet} are used for initialization, and the entire network is then fine-tuned end-to-end on the target dataset during training. The model is trained using the Adam optimizer with an initial learning rate of $5\times 10^{-5}$ and a batch size of $4$ on a single NVIDIA GeForce RTX 3090 GPU.

\subsection{Evaluation Criteria}
We report the performance of all methods in terms of Spearman Rank Correlation Coefficient (SRCC) and Pearson Linear Correlation Coefficient (PLCC). The computational complexity is evaluated using three metrics: latency, number of parameters (Params), and giga multiply-accumulate operations (GMACs). The latency was measured after the model had been sufficiently warmed up to avoid cold-start effects. Specifically, we measured the inference time over 100 test images and computed the average. This measurement includes all preprocessing steps required by the algorithm.

\subsection{Performance Comparison}
The results of SFIQA-Assessor and the compared methods on SFIQA-Bench are summarized in Table~\ref{tab:method_comparison}. From the table, several key observations can be made. First, traditional handcrafted IQA methods such as BRISQUE and NIQE perform poorly on SFIQA-Bench, indicating that NSS-based features are not effective for quantifying the perceptual quality of facial images. Besides, general DNN-based IQA methods such as HyperIQA, TReS, and MANIQA also underperform when compared to perceptual FIQA methods. This is primarily because these models are not specifically optimized for facial features, which limits their ability to capture the nuanced distortions present in face images. 

Second, recognition-oriented FIQA methods exhibit inferior performance compared to perceptual FIQA methods. This highlights the gap between recognition-oriented quality (which focuses on face verification or recognition confidence) and perceptual quality (which aligns with human visual perception). 

Among all evaluated methods, the perceptual FIQA approach DSL-FIQA achieves the best performance in terms of noise, sharpness, contrast, and overall quality. However, its practical utility is limited by its extremely high computational cost---over $200$ million parameters and $2,000$ GFLOPs---which makes it less suitable for real-world deployment. In contrast, our SFIQA-Assessor-S achieves the second-best overall performance, with only a $0.52\%$ drop in SRCC compared to DSL-FIQA. Notably, it does so with over $10\times$ fewer parameters, more than $600\times$ fewer MACs, and approximately $40\times$ lower latency, highlighting its superior balance between effectiveness and efficiency.

When comparing the different variants of SFIQA-Assessor, the performance drops remain minimal. Specifically, SFIQA-Assessor-XS and SFIQA-Assessor-XXS show only a $0.37\%$ and $0.74\%$ reduction for the overall quality dimension, respectively, relative to SFIQA-Assessor-S. However, these reductions come with substantial gains in efficiency. The number of parameters is reduced by $53.5\%$ and $70.3\%$, GMACs are reduced by $45.3\%$ and $62.9\%$, and latency is reduced by $11.8\%$ and $22\%$, respectively, confirming the scalability and deployment-friendliness of the proposed architecture.

The performance on GFIQA-20K and CGFIQA-40K is reported in Table~\ref{tab:GFIQA}. From the results, we observe similar conclusions to those drawn from Table~\ref{tab:method_comparison}, further confirming the effectiveness and efficiency of our proposed methods. In addition, the performance of all methods on GFIQA-20K and CGFIQA-40K is significantly higher than on SFIQA-Bench, suggesting that SFIQA-Bench contain more complex and challenging distortions. This highlights its contribution as a more difficult and comprehensive benchmark for advancing research in face image quality assessment.

\begin{table*}[!t]
  \centering
  \scriptsize
  \caption{
    Cross-dataset performance comparison of representative FIQA methods. The first row indicates the training datasets, and the second row indicates the corresponding testing datasets. SRCC and PLCC are reported for each training–testing pair. For SFIQA-Bench, results are reported using the overall quality dimension. 
  }
  \label{tab:cross_data}
  \renewcommand{\arraystretch}{1}
  \begin{tabular}{l
                  *{2}{cc} 
                  *{2}{cc} 
                  *{2}{cc} 
                  }
    \toprule
    \multirow{3}{*}{Method} 
    & \multicolumn{4}{c}{SFIQA-Bench} 
    & \multicolumn{4}{c}{GFIQA-20K} 
    & \multicolumn{4}{c}{CGFIQA-40K} \\
    \cmidrule(lr){2-5} \cmidrule(lr){6-9} \cmidrule(lr){10-13}
    & \multicolumn{2}{c}{GFIQA-20K} & \multicolumn{2}{c}{CGFIQA-40K}
    & \multicolumn{2}{c}{SFIQA-Bench} & \multicolumn{2}{c}{CGFIQA-40K}
    & \multicolumn{2}{c}{SFIQA-Bench} & \multicolumn{2}{c}{GFIQA-20K} \\
    \cmidrule(lr){2-3} \cmidrule(lr){4-5}
    \cmidrule(lr){6-7} \cmidrule(lr){8-9}
    \cmidrule(lr){10-11} \cmidrule(lr){12-13}
    & SRCC & PLCC & SRCC & PLCC 
    & SRCC & PLCC & SRCC & PLCC 
    & SRCC & PLCC & SRCC & PLCC \\
    \midrule
    DSL-FIQA~\cite{chen2024dsl}
      & 0.7573 & 0.7788 
      & 0.8338 & 0.8263 
      & 0.5511 & 0.5601 
      & 0.9710 & 0.9615 
      & 0.5626 & 0.5802 
      & 0.9491 & 0.9392 \\
    SFIQA-Assessor-S
      & 0.6666 & 0.7092 
      & 0.7499 & 0.7709 
      & 0.5890 & 0.6277 
      & 0.9767 & 0.9750 
      & 0.6029 & 0.6228 
      & 0.9350 & 0.9376 \\
    \bottomrule
  \end{tabular}
\end{table*}

\begin{table}[!t]
\centering
\caption{Performance comparison of different backbone networks. SRCC and PLCC are averaged over the six quality dimensions.}
\label{tab:Backbones}
\resizebox{\linewidth}{!}{
\begin{tabular}{ccccc}
\toprule
Backbones & SRCC  & PLCC  & Params (M)  & GMACs \\
\midrule
EdgeNeXt-S\cite{maaz2022edgenext}        & \textbf{0.9317} & \textbf{0.9510} & \underline{17.76}  & 3.75  \\
MobileNetV4-Conv-M\cite{qin2024mobilenetv4}     & 0.9041  & 0.9306 & 23.42   & 5.67   \\
EfficientFormerV2-S1\cite{li2022efficientformer}       & 0.9150 & 0.9378 & 19.05  & \textbf{2.75} \\
RepViT-M1.0\cite{wang2024repvit}            & 0.9085 & 0.9352 & \textbf{16.10}  & \underline{3.36}  \\
\bottomrule
\end{tabular}
}
\end{table}

\subsection{Cross-dataset Evaluation}

We evaluate two top-performing FIQA methods, DSL-FIQA and our SFIQA-Assessor, in a cross-dataset setting to assess the generalization capability when trained on different datasets. The experimental results are summarized in Table~\ref{tab:cross_data}, where the overall quality dimension is used for evaluation on SFIQA-Bench. From the table, we observe that when models are trained on GFIQA-20K or CGFIQA-40K and evaluated on the other, the performance remains consistently high, with all SRCC values exceeding 0.90. It indicates these two datasets share very similar image distribution. In contrast, when the models are trained on GFIQA-20K or CGFIQA-40K and tested on SFIQA-Bench, the performance drops significantly, further suggesting that SFIQA-Bench contains more complex distortions and diverse content, thus highlighting importance of our proposed benchmark in assessing the quality of surveillance facial images. Interestingly, when trained on SFIQA-Bench and tested on GFIQA-20K or CGFIQA-40K, the models still achieve moderate to strong performance, demonstrating that training on the more challenging SFIQA-Bench enables reasonable generalization to existing FIQA datasets. Finally, it is worth noting that when tested on SFIQA-Bench, our SFIQA-Assessor outperforms DSL-FIQA by a large margin, verifying the effectiveness of our model design for assessing the quality of surveillance facial images.

\begin{table}[!t]
\centering
\caption{Ablation study on the performance of individual modules in the proposed model. ``Original View", ``Face View", and ``Eyes-Mouth View" refer to the original loosely cropped image, the compact face image, the eyes-and-mouth image, respectively. ``CFF" denotes Cross-view Feature Fusion, and ``TSA" denotes Task Self-Attention. SRCC and PLCC are averaged over the six quality dimensions.}
\label{tab:Abl}
\resizebox{\linewidth}{!}{
\begin{tabular}{ccc|cc|cccc}
\toprule 
\multicolumn{3}{c|}{\textbf{Input Views}} & \multicolumn{2}{c|}{\textbf{Fusion \& Decoder}} & \multicolumn{4}{c}{\textbf{Metric}} \\
\cmidrule(lr){1-3} \cmidrule(lr){4-5} \cmidrule(lr){6-9}
\begin{tabular}[c]{@{}c@{}}Original\\ View\end{tabular} & 
\begin{tabular}[c]{@{}c@{}}Face\\ View\end{tabular} & 
\begin{tabular}[c]{@{}c@{}}Eyes-Mouth\\ View\end{tabular} & 
CFF & TSA & SRCC & PLCC & Params (M) & GMACs \\
\midrule
\ding{51} & \ding{55} & \ding{55} & \ding{51} & \ding{51} & 0.9186 & 0.9374 & \textbf{6.91} & \textbf{1.32} \\
\ding{51} & \ding{51} & \ding{55} & \ding{51} & \ding{51} & \underline{0.9279} & 0.9482 & \underline{12.34} & \underline{2.54} \\
\ding{51} & \ding{55} & \ding{51} & \ding{51} & \ding{51} & 0.9219 & 0.9437 & \underline{12.34} & \underline{2.54} \\
\ding{51} & \ding{51} & \ding{51} & \ding{51} & \ding{55} & 0.9093 & 0.9267 & 17.01 & 3.74 \\
\ding{51} & \ding{51} & \ding{51} & \ding{55} & \ding{51} & 0.9265 & \underline{0.9487} & 16.64 & 3.15 \\
\ding{51} & \ding{51} & \ding{51} & \ding{51} & \ding{51} & \textbf{0.9317} & \textbf{0.9510} & 17.76 & 3.75 \\
\bottomrule
\end{tabular}
}
\end{table}

\subsection{Ablation Study}
\label{sec:abl}

In this section, we conduct an ablation study to evaluate the effectiveness of each component of our model. The analysis includes comparisons of different backbone networks, various feature fusion strategies, and the contributions of other individual components.

\subsubsection{Backbones}

We evaluate several widely used efficient backbones, including MobileNetV4-Conv-M~\cite{qin2024mobilenetv4}, EfficientFormerV2-S1~\cite{qin2024mobilenetv4}, and RepViT-M1.0~\cite{wang2024repvit}, which have similar or slightly larger parameter counts compared to EdgeNeXt-S, the default backbone used in this work. The performance comparison is presented in Table~\ref{tab:Backbones}. The results indicate that EdgeNeXt-S achieves the highest average SRCC among all evaluated lightweight backbones, while maintaining low computational cost. This highlights the effectiveness of EdgeNeXt's hybrid CNN-Transformer architecture and its use of adaptive kernel sizes in efficiently extracting facial quality features with linear attention complexity.

\subsubsection{Input Views} 
We evaluate the effectiveness of different input views in Table~\ref{tab:Abl}. Both the compact face view and the eyes-and-mouth view contribute positively to the model’s performance. When either of them is removed, the performance decreases, confirming their effectiveness. Notably, discarding the compact face view leads to a greater performance drop compared to removing the eyes-and-mouth view, indicating that the compact face view is more critical. Nevertheless, the eyes-and-mouth view explicitly provides fine-grained local features, which further enhance the model’s ability to assess perceptual quality when combined with the other views.

\subsubsection{Fusion \& Decoder}
In SFIQA-Assessor, we incorporate a cross-view feature fusion (CFF) module to integrate facial features from different views, and a task-specific self-attention (TSA) module in the decoder to learn task-aware representations. We evaluate the effectiveness of these two modules in Table~\ref{tab:Abl}. The results show that removing either module leads to a noticeable drop in performance, highlighting their importance within the proposed framework. The CFF module facilitates the integration of multi-view features into a unified representation, while the TSA module allows the model to adaptively focus on task-relevant quality information. The best performance is achieved when both modules are used together, confirming their effectiveness in improving feature representation and quality prediction.

\section{Conclusion}

In this paper, we present a comprehensive study on surveillance facial image quality assessment, addressing the unique FIQA challenge posed by real-world surveillance scenarios. Towards this goal, we introduce SFIQA-Bench, a large-scale benchmark with multi-dimensional subjective annotations covering six key dimensions: noise, sharpness, colorfulness, contrast, fidelity, and overall quality. With its diverse surveillance image content and comprehensive evaluation dimensions, SFIQA-Bench can serve as a reliable basis for evaluating FIQA models under real-world surveillance conditions. In addition, we propose SFIQA-Assessor, a lightweight and effective multi-task quality assessment model that leverages cross-view feature fusion and task-guided attention to jointly predict multiple quality dimensions. Extensive experiments demonstrate that our method outperforms existing general-purpose IQA and FIQA approaches. In our study, we observed that the sharpness dimension tends to represent a collection of mixed degradations, rather than a single, isolated clarity-related distortion. For example, motion blur, defocus, and low resolution may all be subsumed under this dimension. In future work, we plan to further disentangle and investigate this dimension in greater depth. We hope our contributions can provide valuable insights and tools for advancing real-world surveillance image quality analysis.

\bibliographystyle{IEEEtran}
\bibliography{sample-base} 

\begin{IEEEbiography}[{\vspace{-0.2in}\includegraphics[height=1.2in,clip,keepaspectratio]{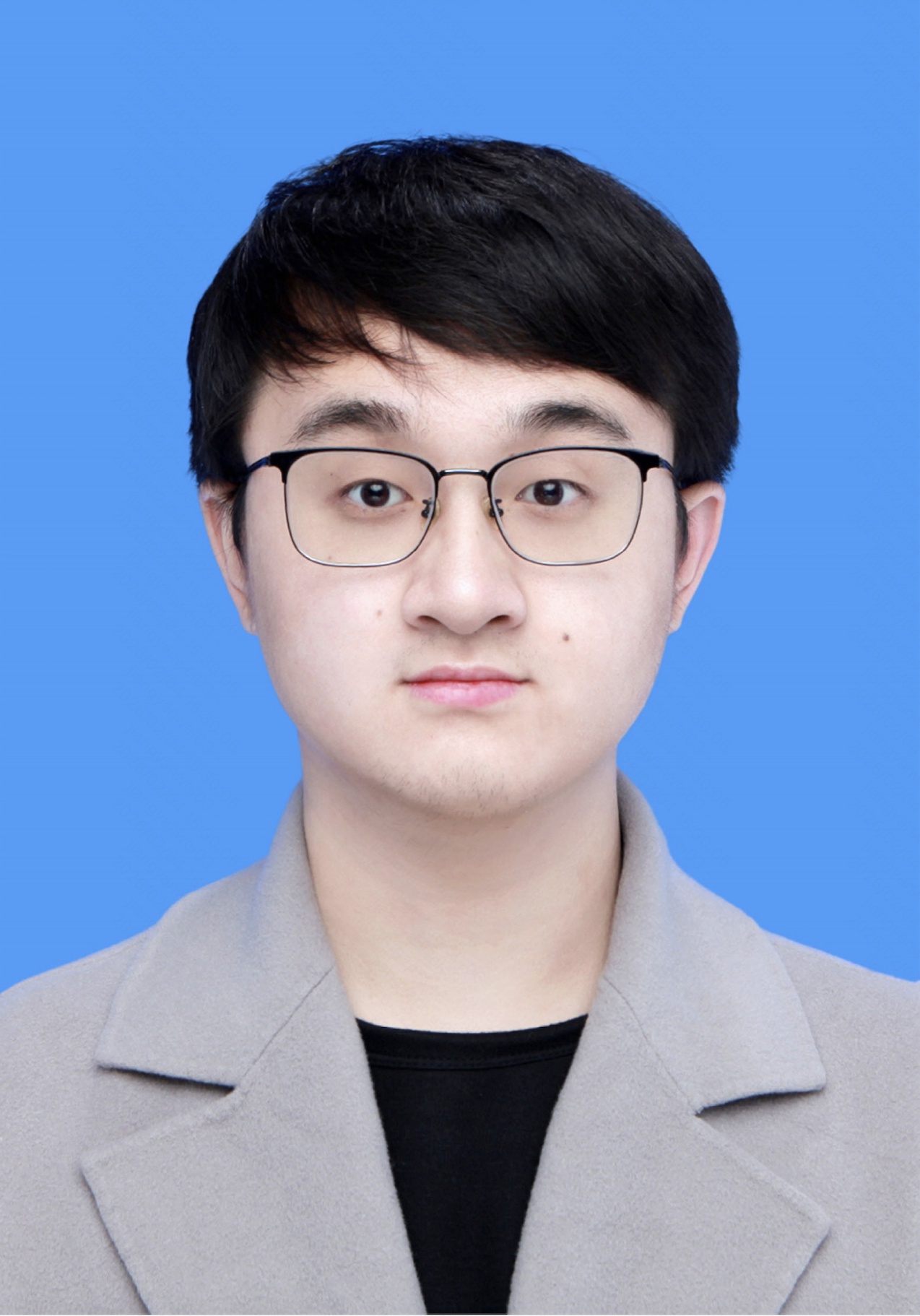}}]{Yanwei Jiang}
received his B.E. degree from Shanghai Jiao Tong University in 2022, where he is currently pursuing the PhD degree with the Institute of Image Communication and Information Processing. His research interests include image quality assessment and multimedia signal processing.
\end{IEEEbiography}
\begin{IEEEbiography}[{\vspace{-0.2in}\includegraphics[height=1.2in,clip,keepaspectratio]{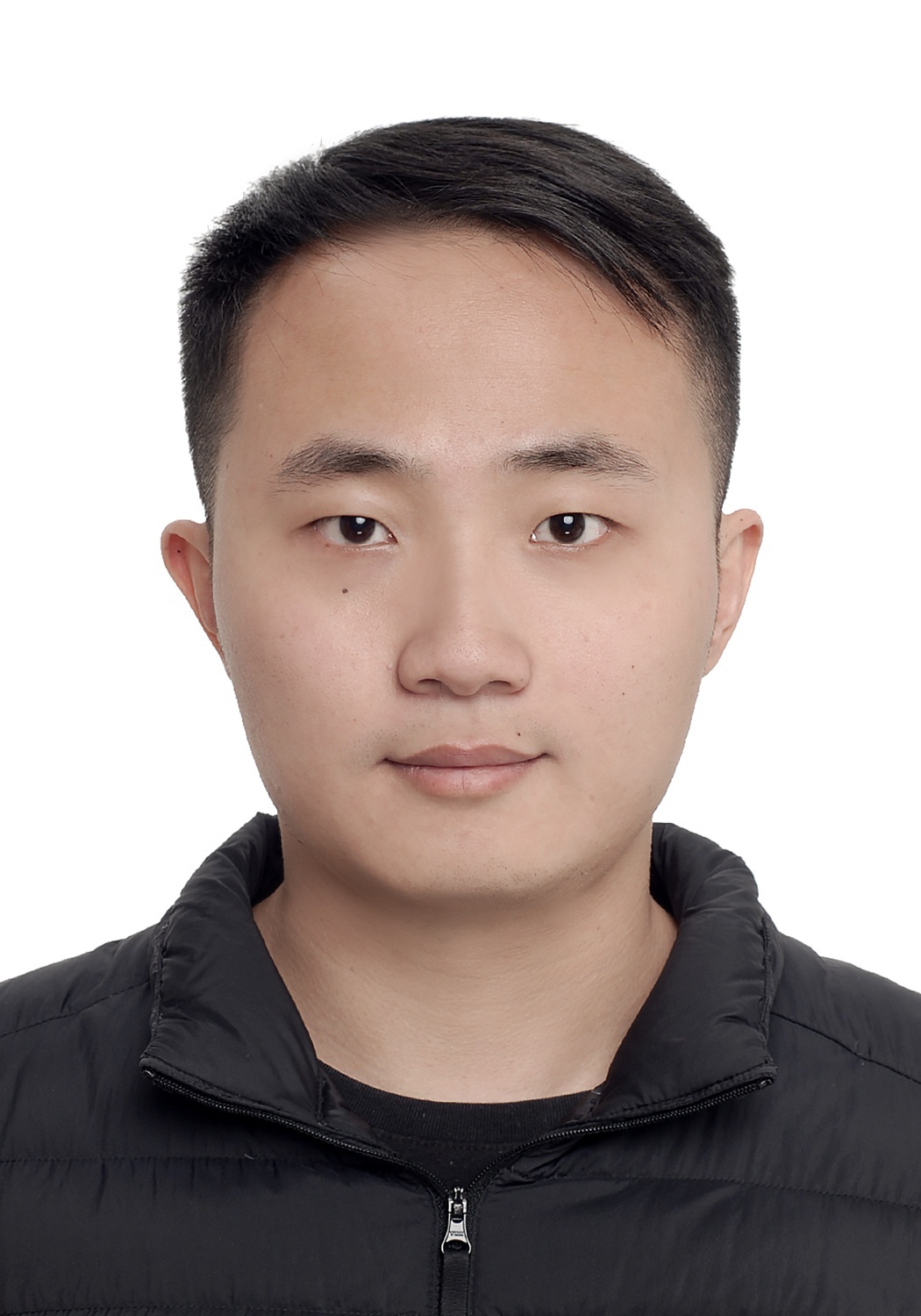}}]{Wei Sun}
received the B.E. degree from the East China University of Science and Technology, Shanghai, China, in 2016, and the Ph.D. degree from Shanghai Jiao Tong University, Shanghai, China, in 2023. From 2023 to 2025, he was a Postdoc at Shanghai Jiao Tong University. He is currently a Research Scientist at East China Normal University. His research interests include image quality assessment, perceptual signal processing, and mobile video processing.
\end{IEEEbiography}

\begin{IEEEbiography}[{\vspace{-0.2in}\includegraphics[height=1.2in,clip,keepaspectratio]{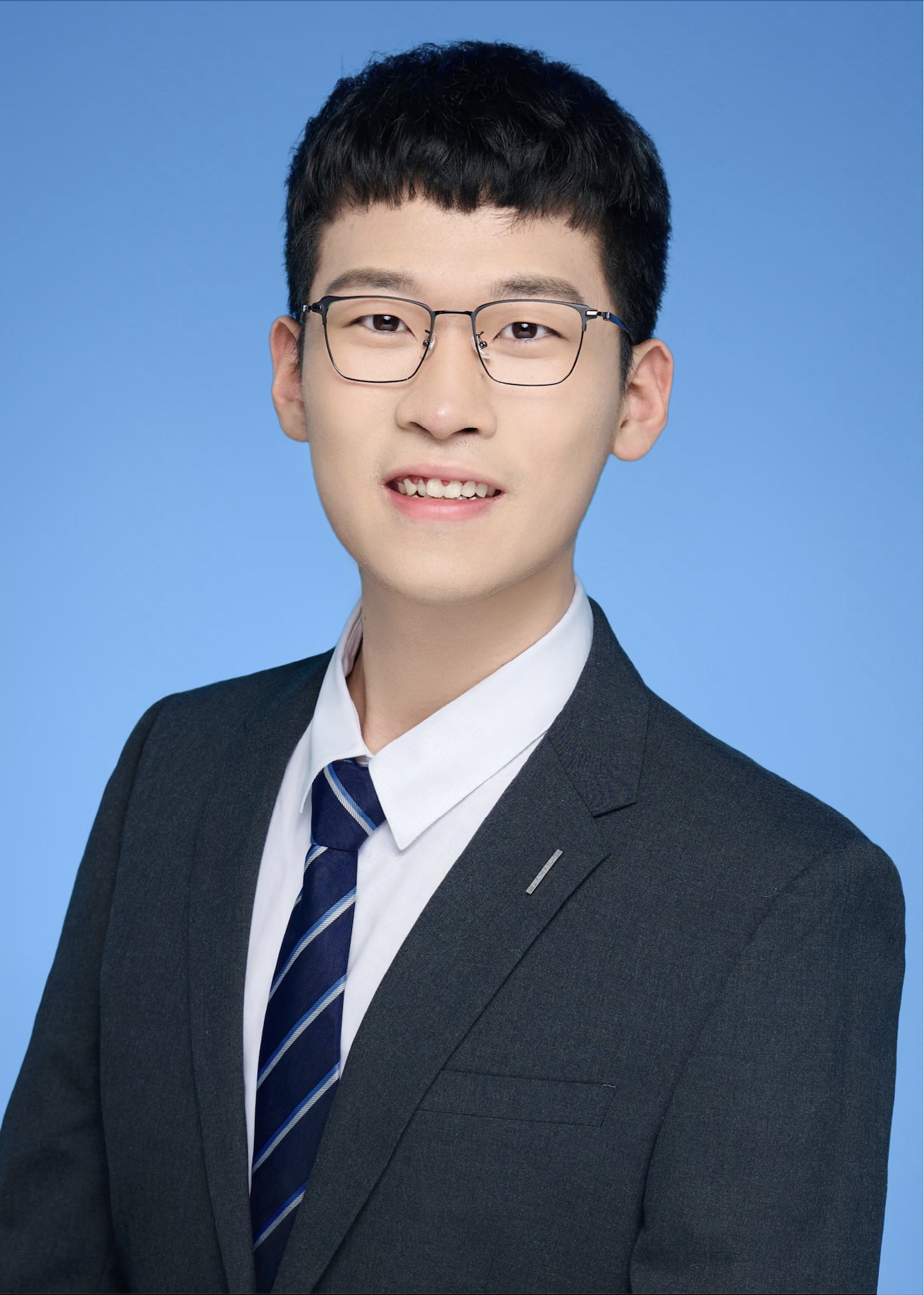}}]{Yingjie Zhou} received his B.E. degree in electronics and information engineering from China University of Mining and Technology in 2023. He is currently pursuing a PhD degree at the Institute of Image Communication and Network Engineering, Shanghai Jiao Tong University, China. His current research interests include digital human quality assessment and sentiment analysis.
\end{IEEEbiography}

\begin{IEEEbiography}[{\vspace{-0.2in}\includegraphics[height=1.2in,clip,keepaspectratio]{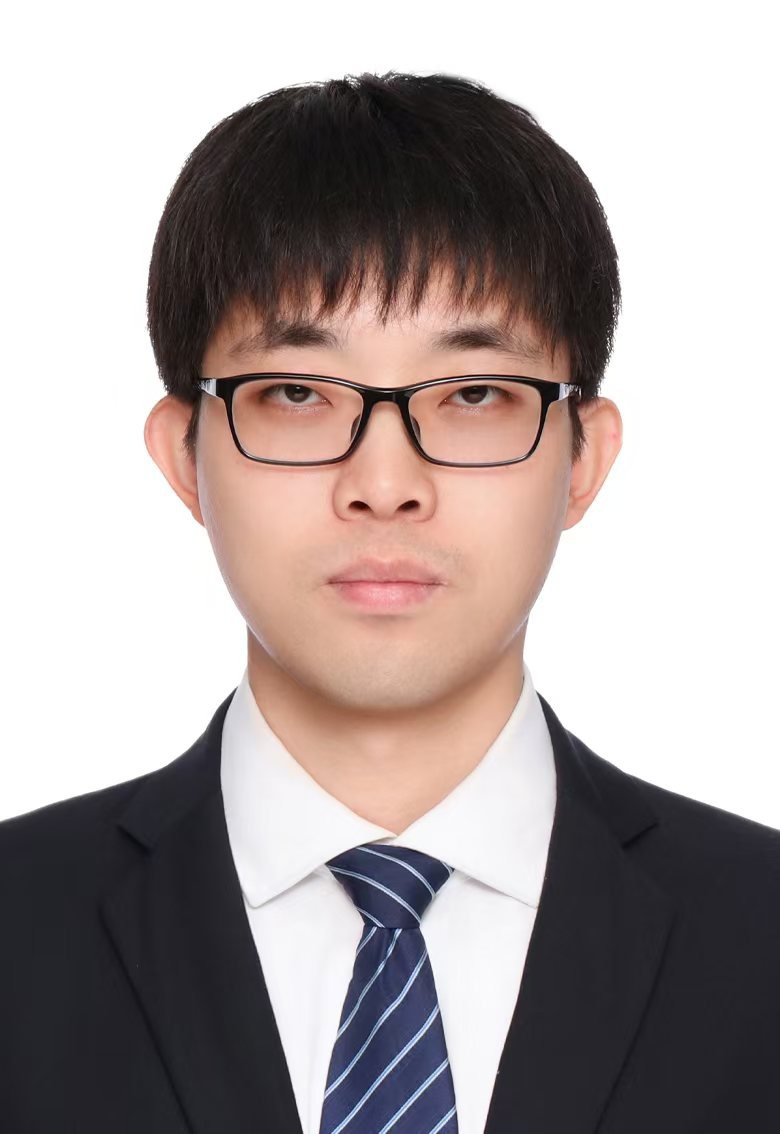}}]{Xiangyang Zhu} received the BE. degree from Huazhong University of Science and Technology, Hubei, China, in 2017, and the Ph.D. degree from City University of Hong Kong, Hong Kong SAR, China, in2024. From 2024, he is a Postdoc at Shanghai AI Lab. His research interests include multimodal representation learning and LLM/MLLM evaluation.
\end{IEEEbiography}

\begin{IEEEbiography}[{\vspace{-0.2in}\includegraphics[height=1.2in,clip,keepaspectratio]{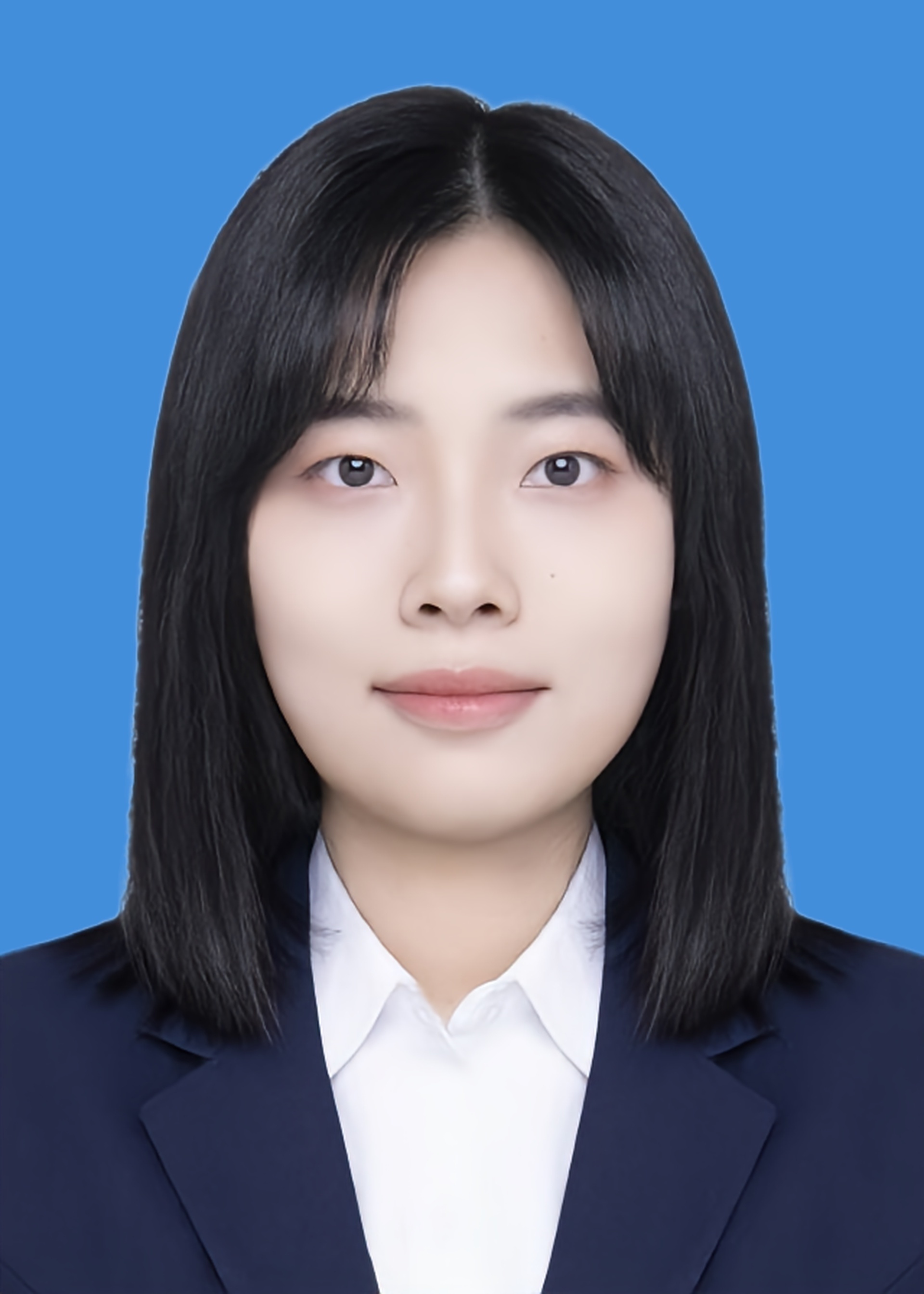}}]{Yuqin Cao} received the B.E. degree from the Shanghai Jiao Tong University, Shanghai, China, in 2021. She is currently working toward a Ph.D. degree with the Institute of Image Communication and Network Engineering, Shanghai Jiao Tong University, Shanghai, China. Her current research interest is in audio-visual quality assessment.
\end{IEEEbiography}

\begin{IEEEbiography}[{\vspace{-0.2in}\includegraphics[height=1.2in,clip,keepaspectratio]{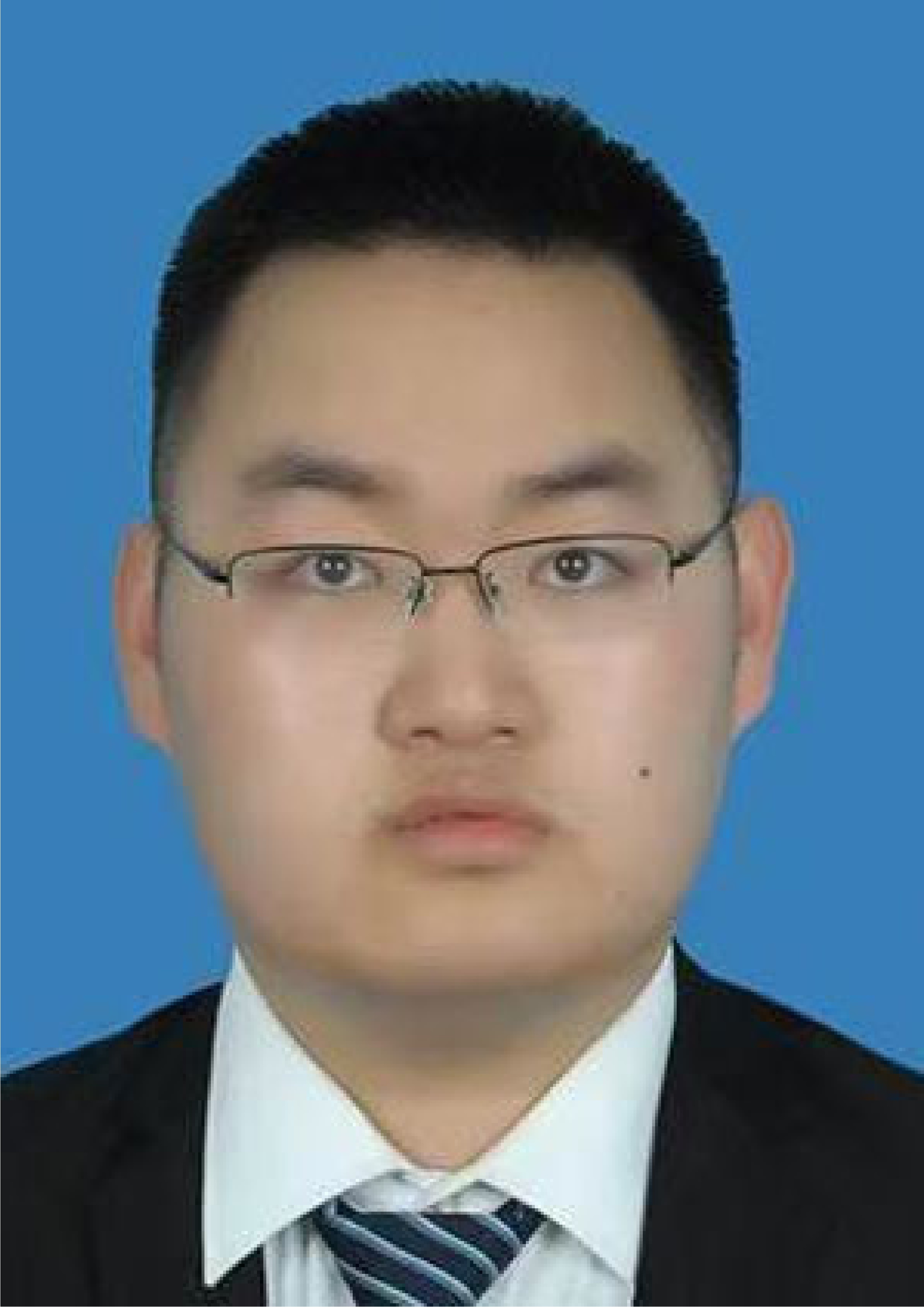}}]{Jun Jia} received the B.S. degree in computer science and technology from Hunan University, Changsha, China, in 2018 and received the PhD degree in the Department of Electronic Engineering, Shanghai Jiao Tong University, Shanghai, China, in 2024. His research interests include computer vision and image processing.
\end{IEEEbiography}

\begin{IEEEbiography}[{\vspace{-0.2in}\includegraphics[height=1.2in,clip,keepaspectratio]{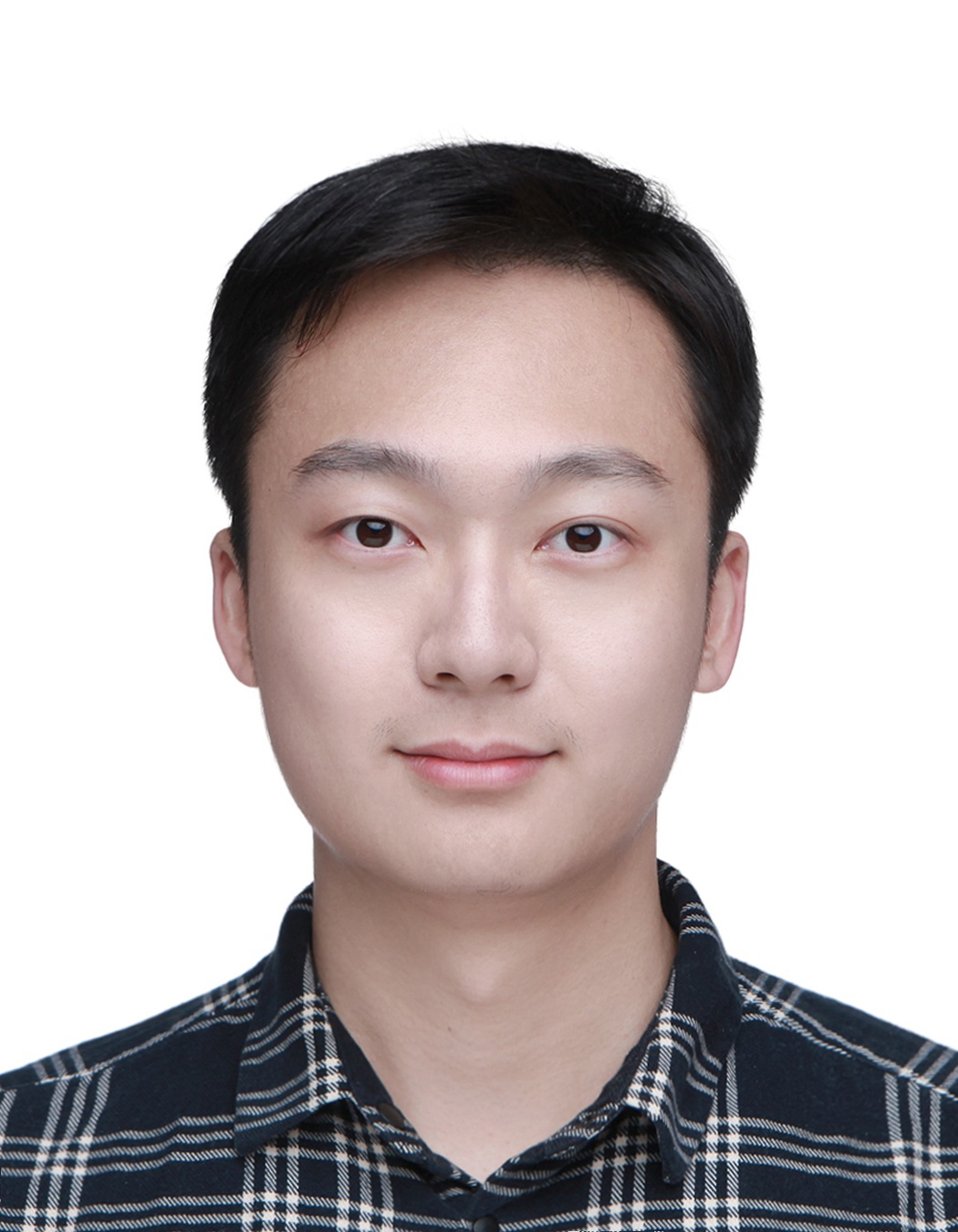}}]{Yunhao Li} received the B.E. degree from Beihang University, Beijing, China, in 2019. He is currently working toward the PhD degree in the Institute of Image Communication and Network Engineering, Shanghai Jiao Tong University, Shanghai, China. His research interests include quality assessment, human-centric understanding and generation.
\end{IEEEbiography}

\begin{IEEEbiography}[{\vspace{-0.2in}\includegraphics[height=1.2in,clip,keepaspectratio]{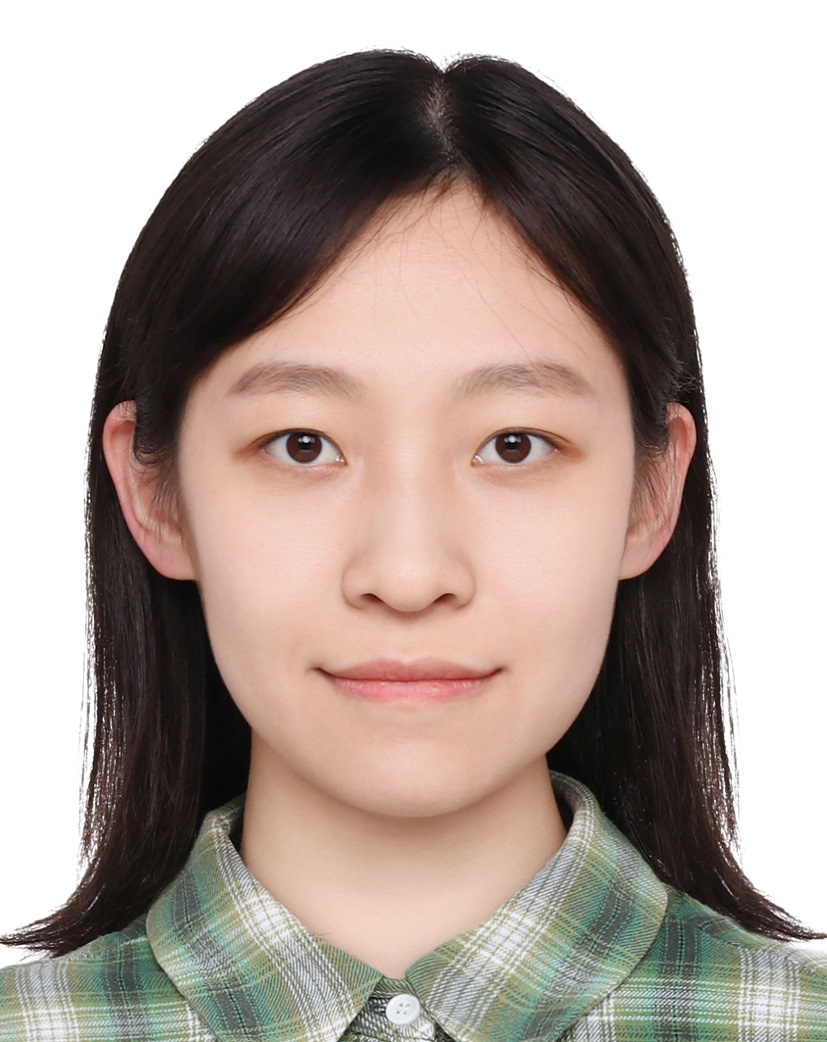}}]{Sijing Wu} received the B.E. degree from Xi'an Jiaotong University, Shaanxi, China, in 2020. She is currently pursuing the Ph.D. degree with the Institute of Image Communication and Network Engineering, Shanghai Jiao Tong University, Shanghai, China. Her research interests include human-centered computer vision, 3D head generation and animation.
\end{IEEEbiography}

\begin{IEEEbiography}[{\vspace{-0.2in}\includegraphics[height=1.2in,clip,keepaspectratio]{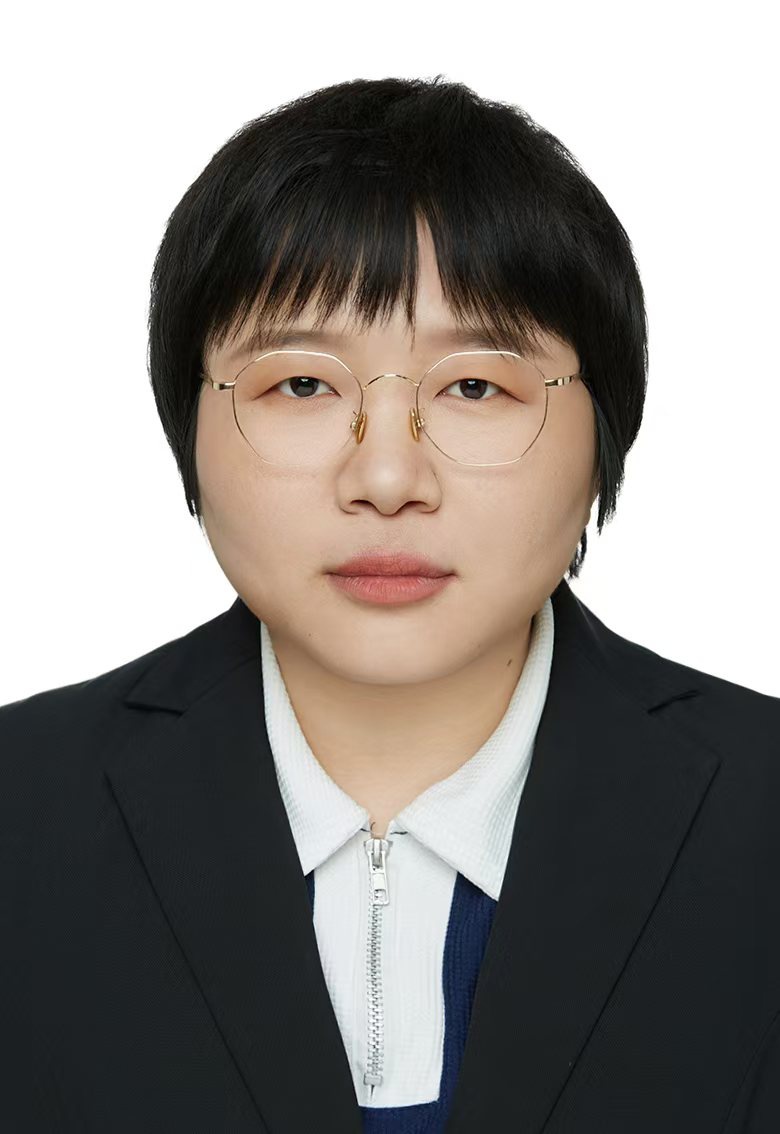}}]{Dandan Zhu} (Member, IEEE) received the Ph.D. degree from Tongji University, Shanghai, China, in 2019. She was a Postdoctoral Researcher with the MoE Key Lab of Artificial Intelligence, Shanghai Jiao Tong University, Shanghai, from 2019 to 2021. She is currently an associate professor with the School of Computer Science and Technology, East China Normal University. Her research interests include multimedia signal processing, computer vision and visual attention modeling.
\end{IEEEbiography}

\begin{IEEEbiography}[{\vspace{-0.2in}\includegraphics[height=1.2in,clip,keepaspectratio]{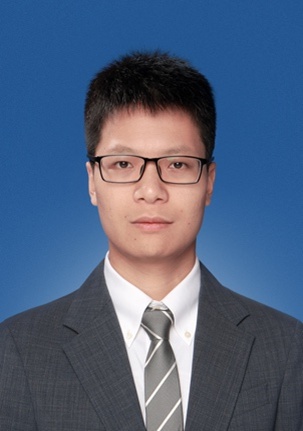}}]{Xiongkuo Min}
received the B.E.
degree from Wuhan University, Wuhan, China, in
2013, and the Ph.D. degree from Shanghai Jiao
Tong University, Shanghai, China, in 2018. He is
currently a tenure-track Associate Professor with
the Institute of Image Communication and Network
Engineering, Shanghai Jiao Tong University. His
research interests include image/video/audio quality
assessment, quality of experience, visual attention
modeling, extended reality, and multimodal signal
processing.
\end{IEEEbiography}

\begin{IEEEbiography}[{\vspace{-0.2in}\includegraphics[height=1.2in,clip,keepaspectratio]{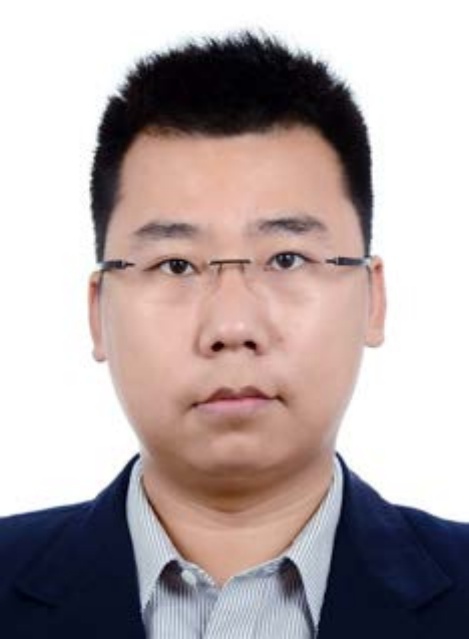}}]{Guangtao Zhai (IEEE Fellow, 2024)} received the B.E. and M.E. degrees from Shandong University, Shandong, China, in 2001 and 2004, respectively, and the PhD degree from Shanghai Jiao Tong University, Shanghai, China, in 2009, where he is currently a Research Professor with the Institute of Image Communication and Information Processing. From 2008 to 2009, he was a Visiting Student with the Department of Electrical and Computer Engineering, McMaster University, Hamilton, ON, Canada, where he was a Post-Doctoral Fellow from 2010 to 2012. From 2012 to 2013, he was a Humboldt Research Fellow with the Institute of Multimedia Communication and Signal Processing, Friedrich Alexander University of Erlangen-Nuremberg, Germany. His research interests include multimedia signal processing and perceptual signal processing.
\end{IEEEbiography}

\end{document}